\documentclass{article}
\usepackage{geometry}
\usepackage[super,compress]{cite}
\usepackage{amsmath}
\usepackage{graphicx}
\usepackage{xcolor}
\usepackage{placeins}
\usepackage{lineno}
\definecolor{posurl}{cmyk}{.9 .9 0 0}
\usepackage[colorlinks=true
,urlcolor=posurl
,anchorcolor=posurl
,citecolor=posurl
,filecolor=posurl
,linkcolor=posurl
,menucolor=posurl
,linktocpage=true
,pdfa=true
]{hyperref}
\begin{document}
\markboth{Vladimir V. Gligorov}{Quark flavour physics: status and future prospects}

%
%

\title{Quark flavour physics: status and future prospects}

\author{Vladimir V. Gligorov\footnote{\href{mailto:vladimir.gligorov@lpnhe.in2p3.fr}{vladimir.gligorov@lpnhe.in2p3.fr}} \\
LPNHE, Sorbonne Universit{\'e}, \\ 
Paris Diderot Sorbonne Paris Cit{\'e}, CNRS/IN2P3}



\maketitle

\begin{abstract}
Quark flavour physics is the study of hadrons, their properties, and their decays into other particles.
As a discipline it simultaneously catalogues the nature of physical states within the Standard Model
of particle physics, and in doing so tests the consistency and completeness of the Standard Model's 
description of reality. Following the discovery of the Higgs field, it is  more essential than
ever to critically examine the Standard Model's own coherence. Precision studies of quark flavour
are one of the most sensitive experimental instruments for this task. I give a brief and necessarily 
selective overview of recent developments in quark flavour physics and discuss prospects for the next 
generation of experiments and facilities, with an emphasis on the energy scales of beyond Standard 
Model physics probed by these types of measurements. 

\end{abstract}



\section{Introduction}	

The Standard Model of particle physics (SM) is the most accurate and complete theory of microscopic
reality. It consists of twelve elementary particles without any internal structure: six quarks and 
six leptons, organised in three families of increasing mass. Four gauge bosons mediate the 
interactions between these elementary SM particles: the photon, $W$, and $Z$ bosons mediate the electroweak 
force, while the gluon mediates the strong force. The Higgs field, whose discovery in 2012 completed the 
Standard Model, is responsible for the rest masses of the elementary particles through its interactions 
with them. Three of the twelve elementary particles: the up and down quarks and the electron, make up 
nearly all matter in the macroscopic world which we inhabit. 

While a wonderfully compact and predictive theory, some of whose predictions have been verified to the
twelfth decimal place\cite{VanDyck:1987ay,Hanneke:2008tm}, the Standard Model is neither a complete theory of 
nature nor is it obviously 
compatible with similarly successful theories of the macroscopic world. Though it describes the mechanism through
which elementary particles acquire mass, as well as the forces which enable these elementary particles to  
combine into complex massive objects, the Standard Model has nothing to say about the gravitational force
between massive objects. The visible matter made of elementary SM particles does not suffice to explain the
masses of galaxies inferred from their motion and the laws of gravity, leading to the postulate that there is
additional ``dark'' matter outside the SM. And most relevantly for this review, the amount of 
matter-antimatter asymmetry allowed in the Standard Model is fundamentally insufficient to explain the 
observed matter-antimatter asymmetry of the universe\cite{Huet:1994jb,Gavela:1994dt}. For all these reasons,
searching for particles and interactions beyond the Standard Model is a key goal of elementary particle 
physics today. 

\begin{figure}[t]
\centering
\includegraphics[width=0.48\linewidth]{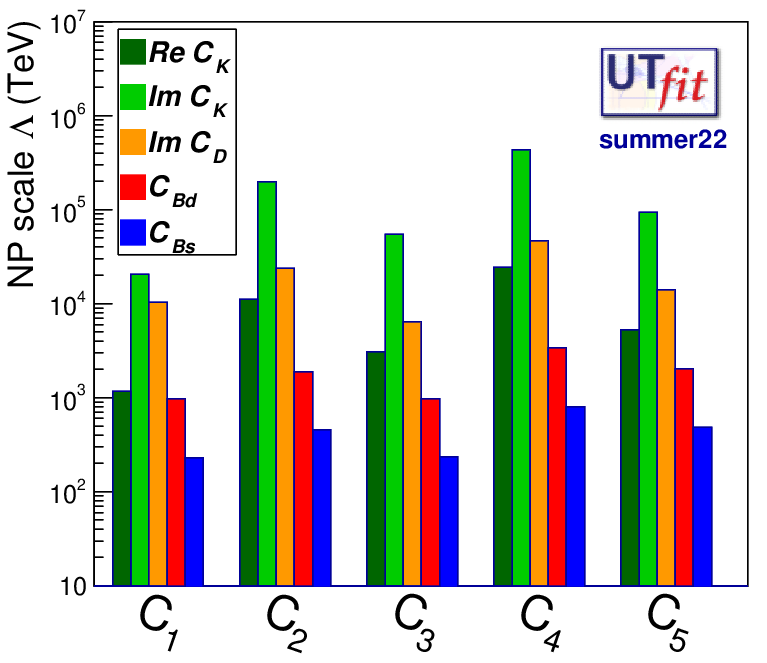}
\includegraphics[width=0.48\linewidth]{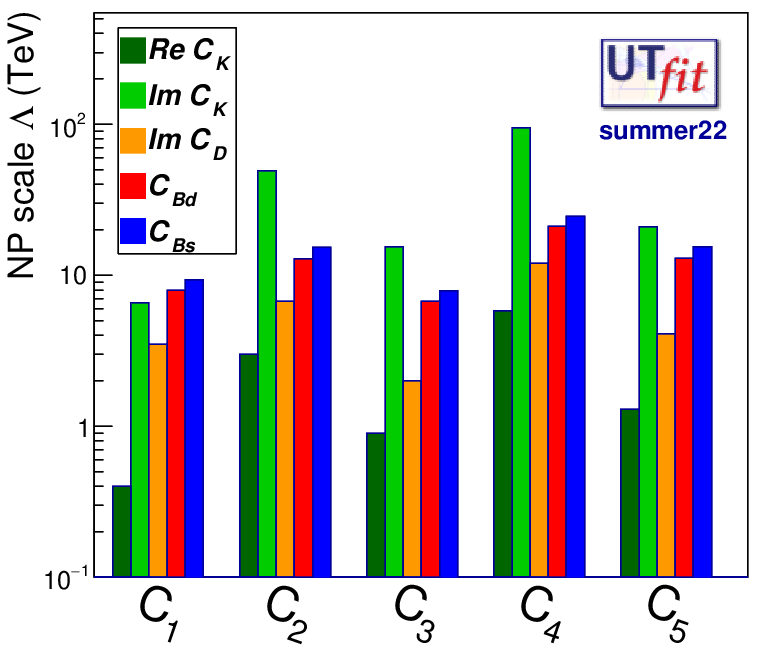}
\caption{Bounds on the energy scale of physics beyond the SM (``NP scale'') arising from different $CP$ 
violating processes. The Wilson coefficients which contribute to the processes in question are labelled 
on the horisontal axis. The left plot shows bounds for tree-level generic couplings, while the right plot 
shows bounds for tree-level MFV couplings. Prepared by the UTFit collaboration\cite{UTfit:2022hsi} and 
used with permission.}
\label{fig:utfitapex}
\end{figure}

Quark flavour physics is the study of hadrons, their properties, and their decays into other particles.  
From a theoretical point of view, the majority of SM parameters describe exactly its flavour
structure, and their origin is one of the fundamental open questions which beyond SM theories aim to answer. 
From an experimental point of view, particles beyond the Standard Model can couple to quarks through 
beyond SM interactions, and modify the properties of hadrons and their decays away from Standard Model 
predictions. The power of such ``indirect'' searches for beyond SM particles is their reach in particle 
mass or, in other words, the energy scale at which physics beyond the SM is to be found. Since beyond 
SM particles can participate virtually in the interations, including decays, of SM hadrons, masses far 
beyond the reach of direct production at today's particle accelerators can be probed. As shown in Figure~\ref{fig:utfitapex}, 
the study of quark flavour has already set extremely stringent limits on certain kinds of beyond SM 
physics, with new particles which couple generically (that is to say, with order one couplings) at 
tree level to SM hadrons ruled out below $10^{4}$--$10^{5}$~TeV. Quark flavour physics therefore gives 
rich information about the nature of new particles long before they can be directly observed, much like 
the observation of beauty meson oscillations in the 1980s led to the conclusion that the top quark must 
be very heavy -- over a decade before it was duly observed at around 170~GeV.

Besides being a tool for precision beyond SM searches, quark flavour physics also offers a unique
window into the workings of Quantum Chromodynamics (QCD). In particular the last decades have seen
an explosion in ``exotic'' hadrons made up of four or five quarks and/or anti-quarks. Numerous QCD models
have been proposed to explain the formation and decay of these four- and five-quark states, ranging from
bound analogues of baryons and mesons (typically referred to as tetraquarks and pentaquarks) to bound 
molecular states made of meson-meson or meson-baryon pairs. The coming decades should allow us to go
from discovery to a precise characterisation of the properties of these exotic hadrons, in turn improving
our understanding of QCD which in turn feeds back into more precise SM predictions for observables 
sensitive to beyond SM effects.

In this brief review I will cover two of the main pillars of today's quark flavour physics: metrology
of the Cabibbo-Kobayashi-Maskawa (CKM) matrix which governs SM quark transitions and searches for beyond 
SM effects in the decays of SM hadrons. As $CP$ violation and mixing in the charm sector\cite{Lenz:2020awd} 
as well as exotic hadron spectroscopy\cite{Gross:2022hyw} have recently been authoritatively reviewed, 
I will not touch on them here. This review is deliberately qualitative, and the reader is invited to consult 
the HFLAV averaging group webpages\cite{HFLAV:2022pwe} for the latest global averages of the relevant experimental 
observables. I will also give an outlook for the next decade of datataking with the current generation 
of experiments, and look ahead to what quantitative and qualitative improvements the next generation of 
experiments and facilities will bring. Natural units are used throughout, and charge conjugation is 
implied unless explicitly stated otherwise.

\section{CKM metrology as a closure test of the SM}

The three by three Cabibbo-Kobayashi-Maskawa (CKM) unitary matrix describes the couplings of quarks to
one another, with SM flavour-changing quark transitions mediated by the weak force. Three real and
one imaginary independent parameters suffice to parametarise the matrix, and it is the non-zero value 
of this imaginary parameter which leads to Charge-Parity ($CP$) violation within the SM. Unitary relations 
between the elements of the CKM matrix define six triangles in the complex plane with equal areas. 
These areas are proportional to the amount of SM $CP$ violation. One of these triangles has sides and 
angles which are particularly convenient to measure using experimentally accessible quark transitions,
in particular transitions which occur in the decay of beauty hadrons. This triangle is commonly referred to as 
``the Unitarity Triangle'', although as we shall see certain crucial tests of the SM CKM mechanism concern
quantities unrelated to this particular triangle. 

If the SM is a complete and self-consistent description of reality, independent measurements of the angles and 
sides of the Unitarity Triangle should be compatible with one another and compatible with a triangle whose 
angles add up to $180^\circ$. Over the past 25 years flavour physics has confirmed that this SM picture of 
flavour-changing quark transitions holds to around the 10\% level\cite{HFLAV:2022pwe,UTfit:2022hsi,Charles:2004jd}. 
As discussed in the introduction, however, new particles and forces must exist at some energy scale. Although 
nothing strictly requires them to mediate flavour-changing quark transitions, most generic models of physics 
beyond the SM do induce such transitions. The self-consistency of the CKM picture of quark transitions 
must consequently break down, and the size of this breakdown in turn gives information about the energy 
scale and flavour structure of whatever lies beyond the SM.

\begin{figure}[t]
\centering
\includegraphics[width=0.48\linewidth]{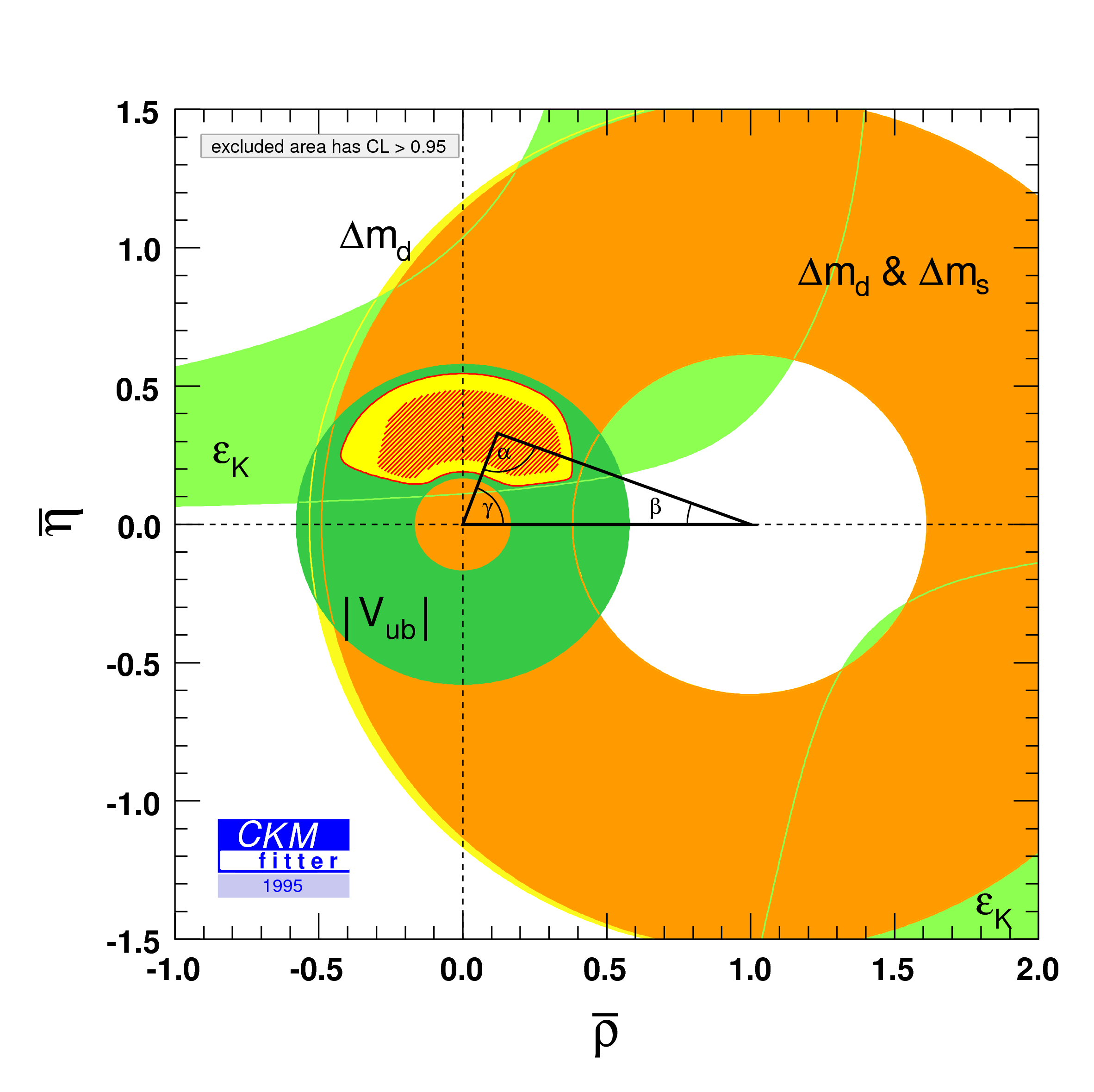}
\includegraphics[width=0.48\linewidth]{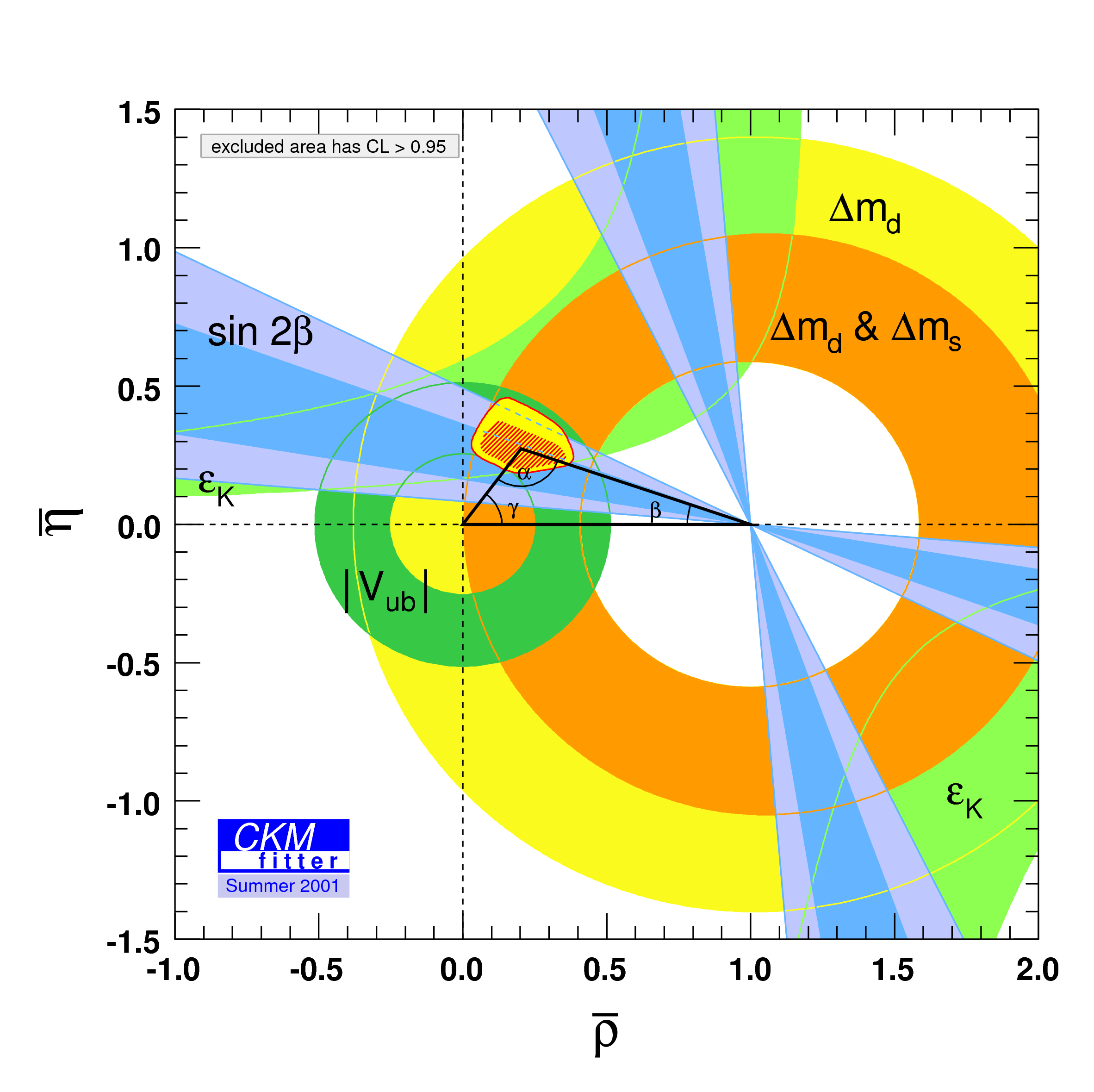}
\includegraphics[width=0.48\linewidth]{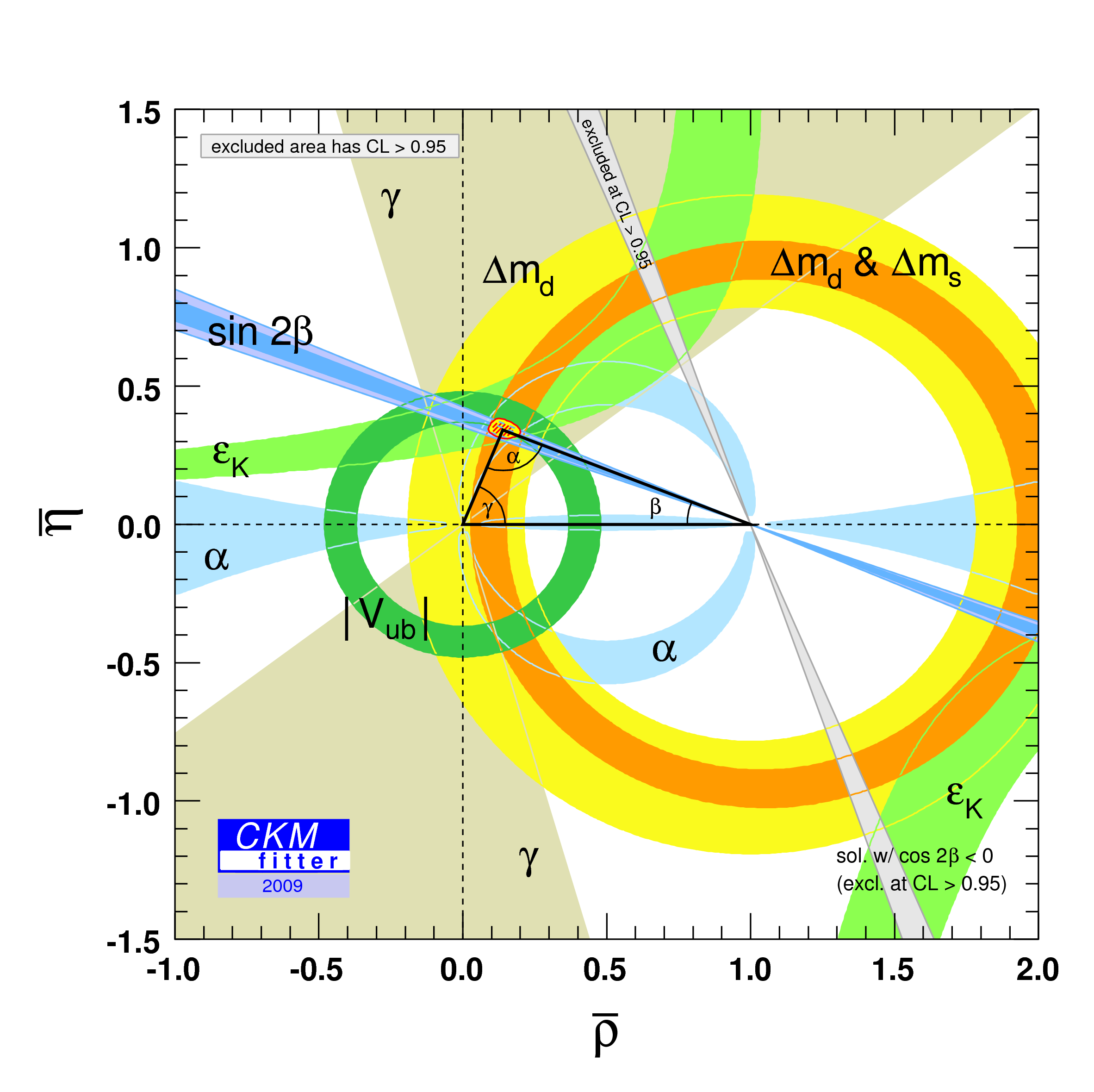}
\includegraphics[width=0.48\linewidth]{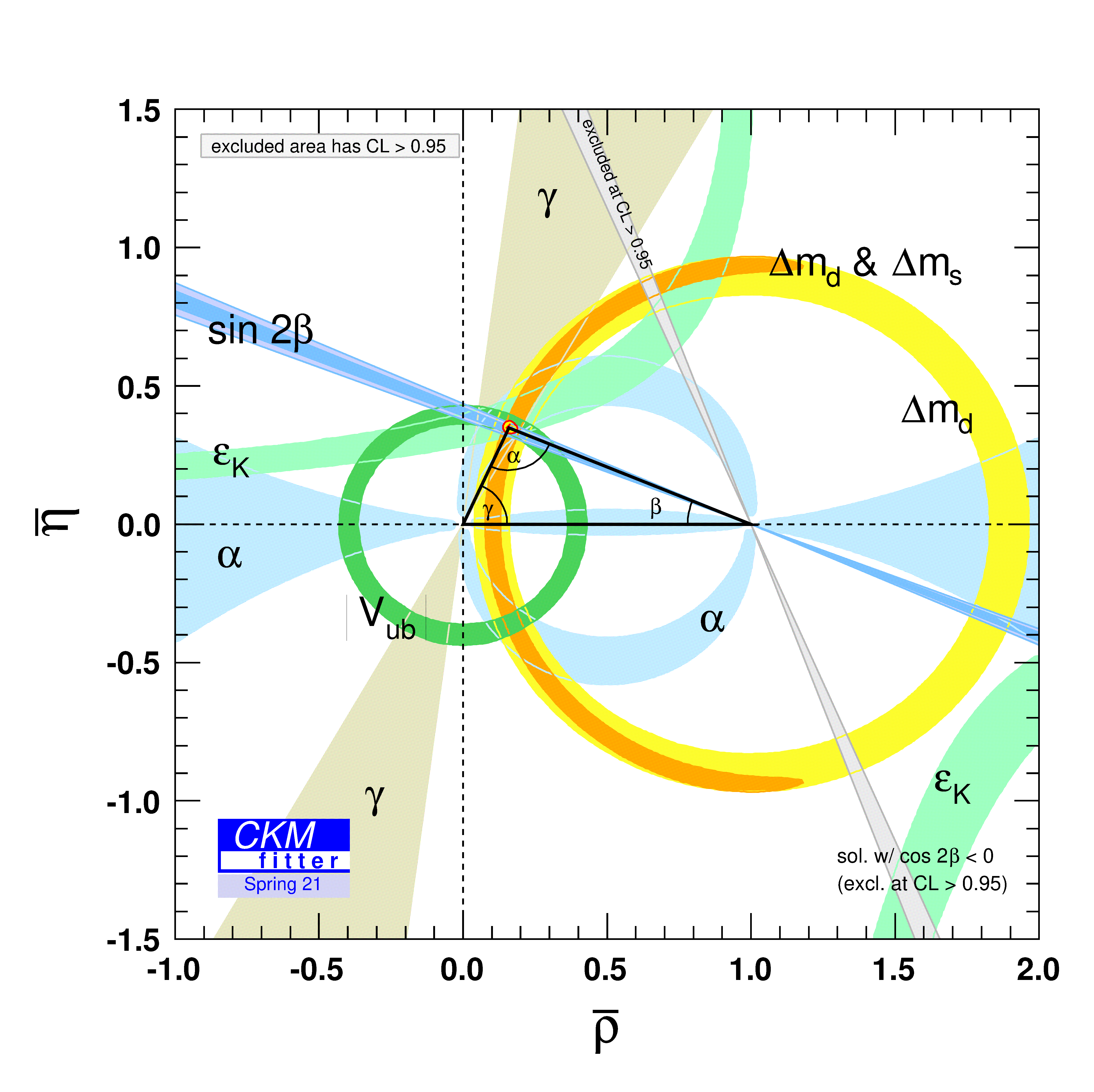}
\caption{Evolution of constraints on the CKM unitarity triangle over time. Clockwise from top left: prior to the 
top quark discovering in 1995; after the first b-factory measurements of the angle $\beta$ in 2001; prior to the 
LHC startup in 2009; in January 2022. Reproduced from CKMFitter\cite{Charles:2004jd}.}
\label{fig:ckmfitterapex}
\end{figure}

Figure~\ref{fig:ckmfitterapex} illustrates the progress made over the past thirty years in the experimental 
determination of the CKM Unitarity Triangle. The axes $\bar{\rho}$ and $\bar{\eta}$ are two of the four parameters
within the Wolfenstein parametrization\cite{Wolfenstein:1983yz} of the CKM matrix. Each of the coloured bands represents the 
constraint in the two-dimensional space of these parameters arising from a particular type of experimental 
measurement. These constraints all meet at a point, typically referred to as the ``apex'' of the unitarity 
triangle, and their consistency is directly related to the closure of the CKM triangle. In what follows
I will focus on those measurements which contribute to the plotted constraints; a comprehensive review of 
direct determinations of individual CKM matrix elements, as well as of $CP$ violation in $K^0_L$ decays which
leads to the light green constraint labelled $\epsilon_{\textrm{K}}$ in Figure~\ref{fig:ckmfitterapex} can be 
found in the PDG\cite{Workman:2022ynf}. 

A particularly popular way to talk about this closure test is by comparing direct and indirect determinations
of the CKM angle $\gamma$, or $\phi_3$ in the Belle experiment notation. The reason is that unlike the other
two angles, $\gamma$ can be directly determined from the relative rates of well-understood tree-level decays of 
beauty hadrons. In addition, all other nuissance parameters which are required to interpret these decay rates in 
terms of $\gamma$ can themselves be experimentally determined without theoretical input. For these reasons the 
total ``theoretical'' uncertainty on the direct determination of $\gamma$ is\cite{Brod:2013sga} roughly 
$\delta\gamma/\gamma \sim 10^{-7}$, far beyond the reach of any concievable experiment. This direct determination
is therefore sometimes referred to as a ``standard candle'' which is then compared to the totality of the other
CKM measurements interpreted in terms of $\gamma$ under the assumption that the CKM triangle is indeed unitary.
There is presently no discrepancy between these determinations, as seen when comparing the latest indirect average 
from CKMFitter\cite{Charles:2004jd} to the latest direct average from HFLAV\cite{HFLAV:2022pwe}

\begin{align}
\gamma_{\textrm{indirect}} &= 65.5^{+1.1}_{-2.7},\\
\gamma_{\textrm{direct}} &= 66.2^{+3.4}_{-3.6}.
\end{align}

\noindent When browsing the literature or hearing talks on this topic it is common to be told\footnote{The author
was guilty of this sin for a long time after graduating, and as is common in such cases now feels particularly
compelled to lecture others on avoiding it.} that $\gamma$ is a standard candle because it is measured using purely 
tree-level decays, and we know that there is no physics beyond the SM in tree-level processes. This is however 
wrong on several counts. For one thing, physics beyond the SM is not universally excluded in tree-level processes 
or for that matter any processes: everything depends on the energy scale of the claimed beyond SM contributions 
and their compatibility with other experimental constraints. In the case of tree-level hadronic decays of beauty 
hadrons such as used to measure $\gamma$, beyond SM contributions of around $10\%$ of the SM rate in fact remain 
allowed\cite{Lenz:2019lvd}. If the direct and indirect determinations of $\gamma$ eventually significantly disagreed, it
would therefore not automatically be the case that beyond SM effects were altering the indirect determination. 

What does remain unique about direct measurements of $\gamma$ is rather their aforementioned experimental and 
phenomenological purity. Since $\gamma$ is a function of the complex CKM matrix element $\textrm{V}_{ub}$, 
measurements of $\gamma$ necessarily involve the observation and measurement of $CP$ violation in various Standard 
Model processes. This in turn requires the interfering decay diagrams to have both a weak (i.e. $\gamma$ itself) 
and a strong (or hadronic) phase difference. The sensitivity to $\gamma$ is maximized when the interfering diagrams 
are of similar size. A good deal of ingenuity has therefore gone into finding pairs of interfering diagrams which 
are as similar to each other in size and understanding their strong phase differences in order to measure $\gamma$.

\begin{figure}[t]
\centering
\includegraphics[width=0.48\linewidth]{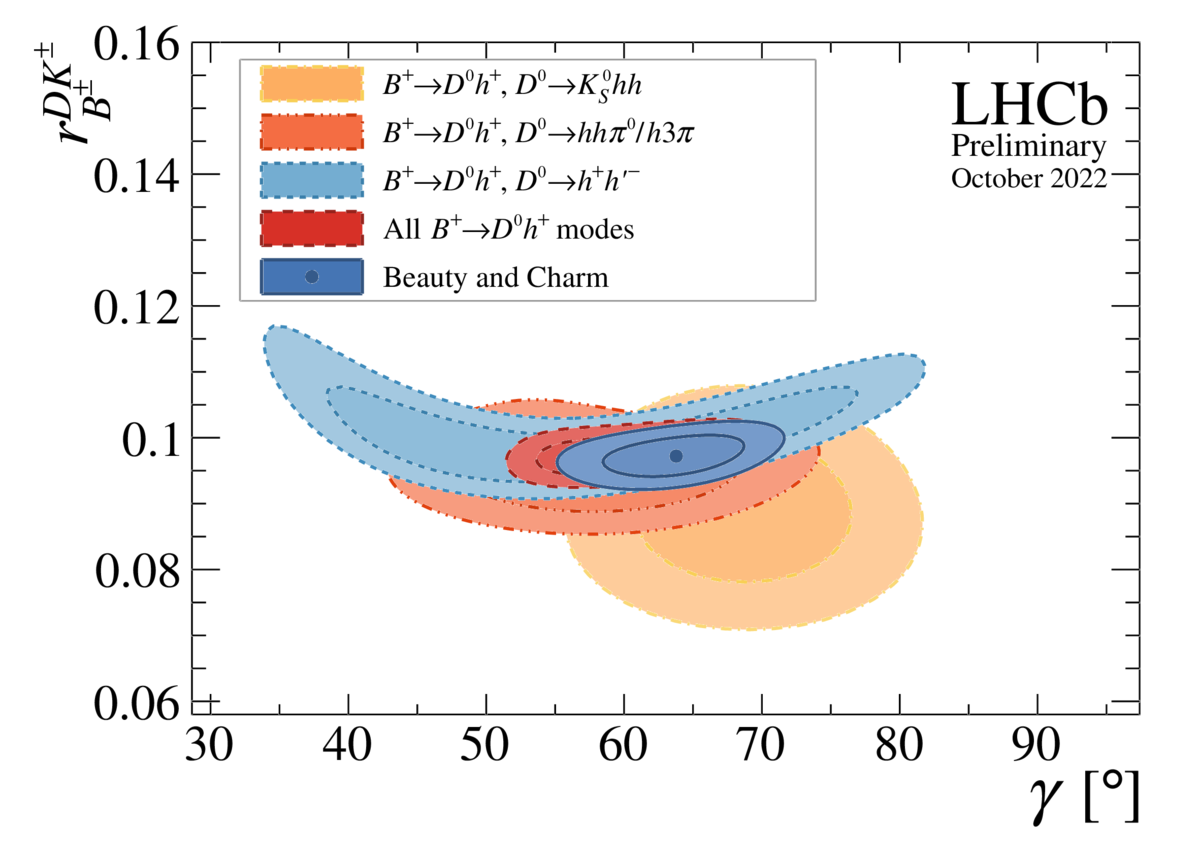}
\includegraphics[width=0.48\linewidth]{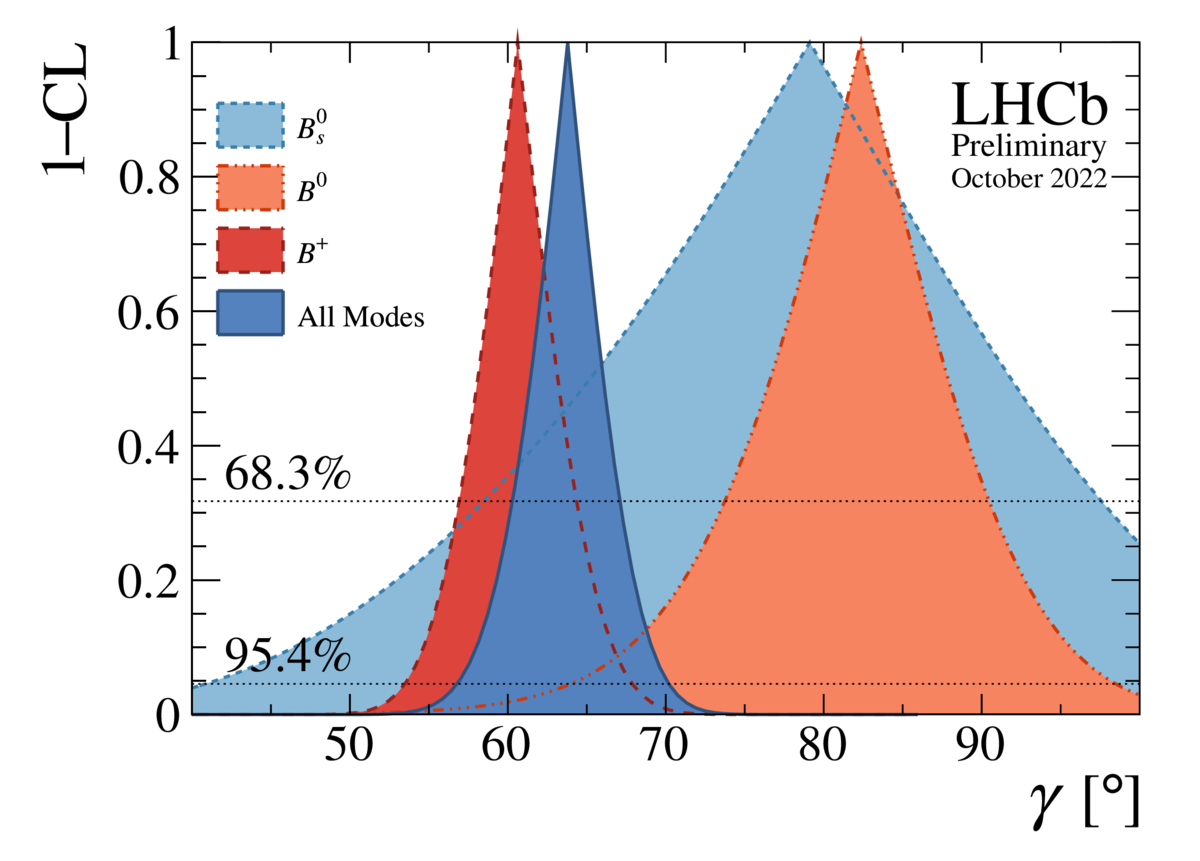}
\caption{(Left) Profile likelihood contours in the two-dimensional space of $r^{DK^\pm}_{B^\pm}$ (the ratio of 
Cabibbo favoured and suppressed $\bar{B}\to D^0 K$ decay diagrams) versus $\gamma$. The contours indicate 
the $68.3\%$ and $95.4\%$ confidence region for each subset of decay modes described in the text and for their 
combination. The impact of including $D^0$ mixing parameters in the global fit is illustrated with a separate
contour. (Right) One dimensional $1-\textrm{CL}$ profiles for $\gamma$ from the combination of 
decay modes labelled in the legend. Both plots show results from LHCb measurements alone. 
Reproduced from LHCb conference note 2022-003\cite{LHCb-CONF-2022-003} which updates the results of LHCb paper 2021-033\cite{LHCb:2021dcr}.}
\label{fig:lhcbgammafit}
\end{figure}

The most sensitive direct determination of $\gamma$ uses $\bar{B}\to D^0 s$ decays in which the $D^0$ meson
can decay into a variety of singly or doubly Cabibbo suppressed final states and $s$ is a hadron state containing 
strangeness. The $B$ hadron can be charged, with the most sensitive final states having $s\equiv K^\pm$, or neutral 
with $s\equiv K^\pm\pi^\mp$. Whether the $B$ hadron is charged or neutral these measurements are performed
integrating over the decay-time distribution of the observed $B$ hadrons. There are so many different processes 
involved, each with their own experimental challenges, that doing them justice would require a dedicated 
review. I will not attempt this here, but rather offer a few general observations about how the different 
measurements fit together. While BaBar\cite{BaBar:2008inr} and Belle\cite{Belle:2021efh} have both measured $\gamma$, 
and Belle~2 will eventually have excellent sensitivity to it, the current world average is almost entirely dominated 
by the average of LHCb measurements\cite{LHCb:2022awq} which is summarized in Figure~\ref{fig:lhcbgammafit}.

The experimentally simplest measurements involve the so-called ``ADS/GLW'' 
family\cite{Gronau:1990ra,Gronau:1991dp,Atwood:1996ci,Atwood:2000ck} of fully charged 
two-body decays $D^0\to K^+K^-, \pi^+\pi^-, K^\pm\pi^\mp$. These are generally measured together in order to 
minimize systematic uncertainties, given that the only real experimental difference between them are the 
signal-to-background ratios. As can be seen in Figure~\ref{fig:lhcbgammafit} these measurements give the most precise 
determination of $r^{DK^\pm}_{B^\pm}$, the ratio of Cabibbo favoured and suppressed $\bar{B}\to D^0 K$ 
decay diagrams, but they do not themselves precisely isolate the value of $\gamma$. For this it is
essential to use multi-body decays such as $D^0\to K^0_\textrm{S} hh, h3\pi$ where $h$ is a charged
pion or kaon.\footnote{The $D^0\to K^0_\textrm{S} h^+h^-$ decays, which were proposed first in this context,
are generally known as BPGGSZ measurements after their originators, but the four-body charged decays follow
exactly the same underlying physics logic.} These decays can proceed through multiple intermediate hadronic 
resonances, in effect increasing the number of potential decay paths which can interfere and thereby increasing 
the sensitivity to $CP$ violation and to $\gamma$ itself. This is also seen on Figure~\ref{fig:lhcbgammafit} 
where these decays isolate much narrower regions in $\gamma$ than the ADS/GLW modes.

The downside of this complexity is that one generally then needs\cite{BondarGamma,Giri:2003ty,Belle:2004bbr,Ceccucci:2020cim} 
either a model of these intermediate resonances which comes with an irreducible and hard to quantify systematic 
uncertainty, or at least knowledge of how the strong phase difference between the decay diagrams varies as
a function of the Dalitz plot of the decay. These are commonly referred to as model-dependent and
model-independent determinations, respectively, and it is the model-independent determinations which drive
the overall sensitivity to $\gamma$ today. The strong phase differences are measured using quantum-correlated
$D^0-\bar{D^0}$ pairs\cite{Bondar:2008hh} at CLEO\cite{CLEO:2010iul} and BESIII\cite{BESIII:2020hlg}. 
Since CLEO is no longer operational the ability of future experiments to measure $\gamma$ with better than 
$1^\circ$ sensitivity crucially depends\cite{BESIII:2020khq,BESIII:2021eud} on 
improved measurements of these strong phase differences in future dedicated runs of BESIII. 

Other time-integrated measurements of $\gamma$ are variations on this same theme, whether they involve
excited charm mesons or beauty baryons. In the long-term it may well be the case that the
so-called double-Dalitz analysis\cite{Gershon:2009qc,Poluektov:2023ucp} of $\bar{B^0}\to D^0(K^0_\textrm{S} h^+h^-, h3\pi) K\pi$ 
decays gives the single most precise measurement of $\gamma$, since it maximizes the available number of potentially interfering 
decay paths. It also however requires particularly large data samples to experimentally constrain all the 
associated nuissance parameters. In addition the Cabibbo-favoured family of $\bar{B}\to D^0 \pi$ decays 
is approaching the point where they will become experimentally sensitive to $CP$ violation, and thereby also 
$\gamma$. Indeed, the entire ensemble of time-integrated $\gamma$ measurements is now sensitive to effects
from $D^0$ meson mixing, such that a combined fit of charm mixing and $\gamma$ observables\cite{LHCb:2022awq} is
required to obtain the best sensitivity to both.

It is also possible to measure $\gamma$ using time-dependent decays of $B^0$ or $B^0_s$ mesons. In this case
the mixing phases $\beta_{(s)}$ enter the determination as an additional nuissance parameter which must be
determined from other measurements. We can then choose whether to constrain the mixing phase in the fit
and measure $\gamma$ or whether to constrain $\gamma$ in the fit and measure the mixing phase. The first
such measurements were performed using $B^0\to D^{(*)\pm} \pi^\mp$ decays\cite{Long:2003wq,Baak:2007gp}, however they have a
poor sensitivity to $\gamma$ because the interfering diagrams are so different in size that their relative
size is too small to be experimentally constrained with any plausibly collectible dataset\cite{Fleischer:2003yb}. It
must therefore be taken from theoretical estimates which come with unacceptably large systematic
uncertainties. The related $B^0_s\to D^\pm_s K^\mp$ decays have no such problem, as the interfering diagrams
are similar in size and all nuissance parameters can be determined from experimental measurements. They
are therefore competitively sensitive to $\gamma$ per unit luminosity at LHCb, and can be used
to determine $\gamma$ with only a twofold trigonometric ambiguity, similarly to the BPGGSZ method.

However what is unique about $B^0_s\to D^\pm_s K^\mp$ is that it is the only pure tree-level process sensitive
to the $B^0_s$ mixing phase $\beta_{s}$. Today the average of direct measurements of $\gamma$ leads
to a roughly $4^\circ$ uncertainty, while the average\cite{HFLAV:2022pwe} of direct measurements of $\beta_{s}$ leads\footnote{As
we will discuss shortly, these measurements actually measure a weak mixing phase which is very nearly equal to
$\beta_{s}$, but is modified in the Standard Model by subleading loop-level diagrams. The size of these subleading
SM contributions is such that ignoring them is certainly correct when using the measured weak phase as input to a
measurement of $\gamma$, but may not be when interpreting it as a test of the Standard Model.} to a roughly
$1^\circ$ uncertainty. So while it is formally possible to interpret the experimental $CP$ violation observables 
in $B^0_s\to D^\pm_s K^\mp$ in terms of either $\gamma$ or $\beta_{s}$, in practice only the interpretation in
terms of $\gamma$ is useful. Since Belle~2 can only measure $\gamma$, and since LHCb's upgrades
and their all-software trigger will primarily improve sensitivity for hadronic final states, this gap in
sensitivity between other direct measurements of $\gamma$ and $\beta_{s}$ will likely narrow in the future. 
Consequently $B^0_s\to D^\pm_s K^\mp$ may in the long run be interpreted as a measurement of $\beta_{s}$.
A similar phenomenology is present in $B^0_{s}\to K^+K^-$ and $B^0\to \pi^+\pi^-$ decays, where $h$ is a pion 
or kaon, which are however loop-level processes in the Standard Model. In addition since the final state of
these decays is a $CP$ eigenstate, there are fewer observables compared to $B^0_s\to D^\pm_s K^\mp$, so it
is necessary to use U-spin symmetry\cite{Fleischer:1999pa} to relate the strong phases in the $B^0$ and $B^0_{s}$ decays in
order to determine\cite{LHCb:2014nbd} $\gamma$ and $\beta_{s}$.

\begin{figure}[t]
\centering
\includegraphics[width=0.48\linewidth]{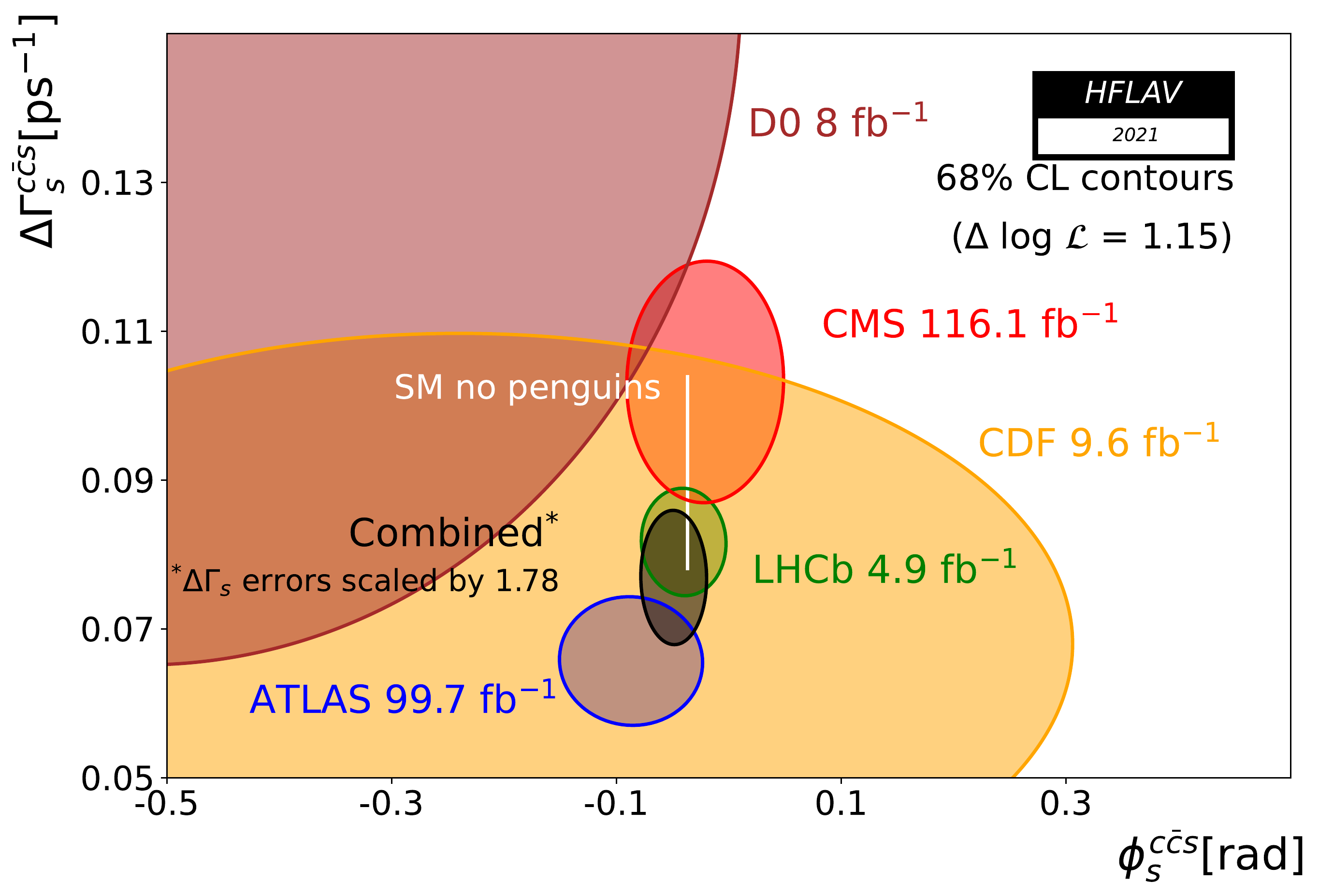}
\includegraphics[width=0.48\linewidth]{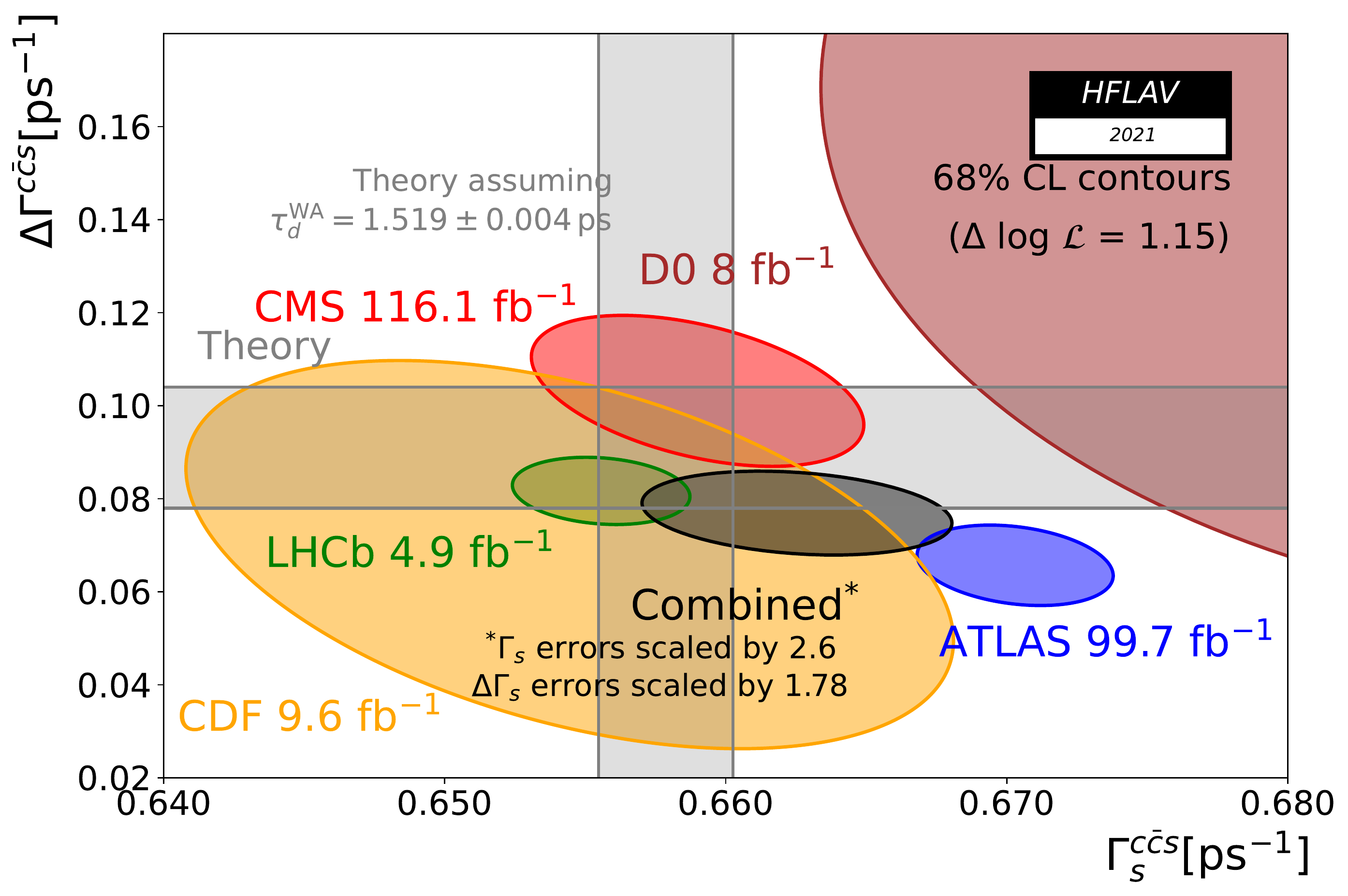}
\caption{Constraints on lifetime and $CP$ violation observables from a combination of $B^0_s\to c\bar{c}s$ processes. 
The plotted contours correspond to $68.3\%$ confidence regions. (Left) the $CP$ violating phase $\phi_s$ versus the
width splitting $\Delta\Gamma_s$. (Right) The effective $B^0_s$ lifetime $\Gamma_s$ versus $\Delta\Gamma_s$.  
Reproduced from HFLAV\cite{HFLAV:2022pwe}.}
\label{fig:phisdgs}
\end{figure}

Measurements of $\beta_{s}$ are themselves of great interest as a test of the Standard Model. This phase, 
equal to $-2\texttt{arg}(\textrm{V}_{ts}\textrm{V}^*_{tb}/\textrm{V}_{cs}\textrm{V}^*_{cb})$, is not directly related 
to any angle or side of the Unitarity Triangle illustrated in Figure~\ref{fig:ckmfitterapex}. Since all of its 
terms are real to leading order in the Wolfenstein parametrisation we can immediately intuit that $\beta_{s}$ should 
be very small in the SM, while it can be arbitrarily enhanced in beyond SM scenarios. The current indirect 
prediction\cite{Charles:2004jd} is $\beta_{s} = -0.0368^{+0.0009}_{-0.0007}$. Most measurements of $\beta_{s}$ are based 
on dominantly tree-level $b\to c\bar{c}s$ processes, which are nevertheless affected by subleading loop-level diagrams. 
This means the experimentally observable phase $\phi_s$ is not exactly the same as $\beta_{s}$. 

The importance of these loop-level diagrams 
can be experimentally constrained\cite{Faller:2008gt,Fleischer:1999zi} by studying related $B^0$ decay processes and relating one to the other 
using $SU(3)$ symmetry. Such analyses\cite{LHCb:2014xpr} have generally affirmed that the subleading penguin diagrams are indeed
small, as expected. The average of the weak phase measured in $b\to c\bar{c}s$ processes is $\phi_s = -0.049\pm 0.019$, 
in excellent agreement with the SM value of $\beta_{s}$, as illustrated in Figure~\ref{fig:phisdgs}. It is however interesting
to notice that while the determination of $\phi_s$ agrees well between experiments, measurements of the $B^0_s$ 
effective lifetime and of the width-difference between the two $B^0_s$ mass eigenstates are in some tension with 
each other. If that tension persists it may introduce limiting systematic uncertainties in other time-dependent analyses 
of $B^0_s$ decays which rely on them as input, including $B^0_s\to D^\pm_s K^\mp$. Resolving this tension is therefore
of the utmost importance. Measurements of the mixing phase have also been carried out\cite{LHCb:2023exl} in the dominantly 
loop-level $b\to s\bar{s}s$ and $b\to d\bar{d}s$ processes, and agree well with SM predictions and with the
value of $\phi_s$ measured in $b\to c\bar{c}s$ processes.

Measurements of the $B^0_d$ and $B^0_s$ meson oscillation frequencies can be interpreted in terms of 
constraints in the $\bar{\eta},\bar{\rho}$ plane, although with non-negligible hadronic uncertainties\cite{DiLuzio:2019jyq}.
Current experimental measurements are in good agreement with other determinations of the Unitarity Triangle
apex. The CKM angle $\beta$, or $\phi_1$ in the Belle experiment notation, was the first of the CKM angles
to be experimentally measured. 
It is equivalent to $-2\texttt{arg}(\textrm{V}_{cd}\textrm{V}^*_{cb}/\textrm{V}_{td}\textrm{V}^*_{tb})$ and 
since $\textrm{V}_{td}$ is complex while the numerator and denominator have similar magnitudes, $\beta$ is expected 
to be large in the SM. The original determinations of $\beta$ by BaBar\cite{BaBar:2001pki} and Belle\cite{Belle:2001zzw} through the
observation of time-dependent $CP$ violation in $B^0\to J/\psi K^0$ are generally considered the moment when 
the CKM picture of quark mixing was confirmed to be correct at first order. Similarly to $\beta_{s}$, 
$\beta$ is predominantly measured using $b\to c\bar{c}s$ transitions, interpreting the experimental observables
under the assumption that the subleading loop-level decay diagrams are negligible. The last of the CKM angles, 
$\alpha$ or $\phi_2$, must be determined through a combined analysis\cite{Gronau:1990ka,Gronau:2005pq} of 
various $b\to u\bar{u}d$ processes. There is no path to $\alpha$ which does not involve at least some, and often 
multiple, $\pi^0$ mesons in the final state, which makes it much more difficult to measure at LHCb than at BaBar 
and Belle. There has therefore been relatively little progress on $\alpha$ in the last decade. The 
measurements of both $\alpha$ and $\beta$ remain limited by the statistical sensitivity of current datasets and 
their current values agree well with other determinations of the Unitarity Triangle apex. 

The final constraint visible in Figure~\ref{fig:ckmfitterapex} comes from measurements of the CKM matrix element 
$\textrm{V}_{ub}$, which is in practice normalised by the element $\textrm{V}_{cb}$ in order to 
cancel certain hadronic uncertainties. An equivalent constraint could be obtained using direct measurements
of $\textrm{V}_{td}$, however no such measurements currently exist. Both $\textrm{V}_{ub}$ and 
$\textrm{V}_{cb}$ are determined from measurements of time-integrated $B^{0,+} \to (u,c) X$ decay rates. 
The essential difficulty, and main source of uncertainty, in these measurements are the hadronic form factors 
and decay constants which must be externally constrained in order to interpret the decay rates in terms of
the CKM matrix parameters. The experimental choice is between isolating specific exclusive processes such as 
$B^0 \to \pi^- l^+ \nu_l$ and performing an inclusive measurement\footnote{Inclusive measurements are only possible 
at $e^+e^-$ colliders because they rely on the well-known initial state in order to infer the presence and kinematics
of a $b\to (u,c)$ transition.} by tagging the presence of a $b\to (u,c)$ transition using global event properties. 
The tradeoff is that exclusive processes have much smaller backgrounds, while inclusive processes require fewer
external inputs and hence have smaller theoretical uncertainties.

\begin{figure}[t]
\centering
\includegraphics[width=0.7\linewidth]{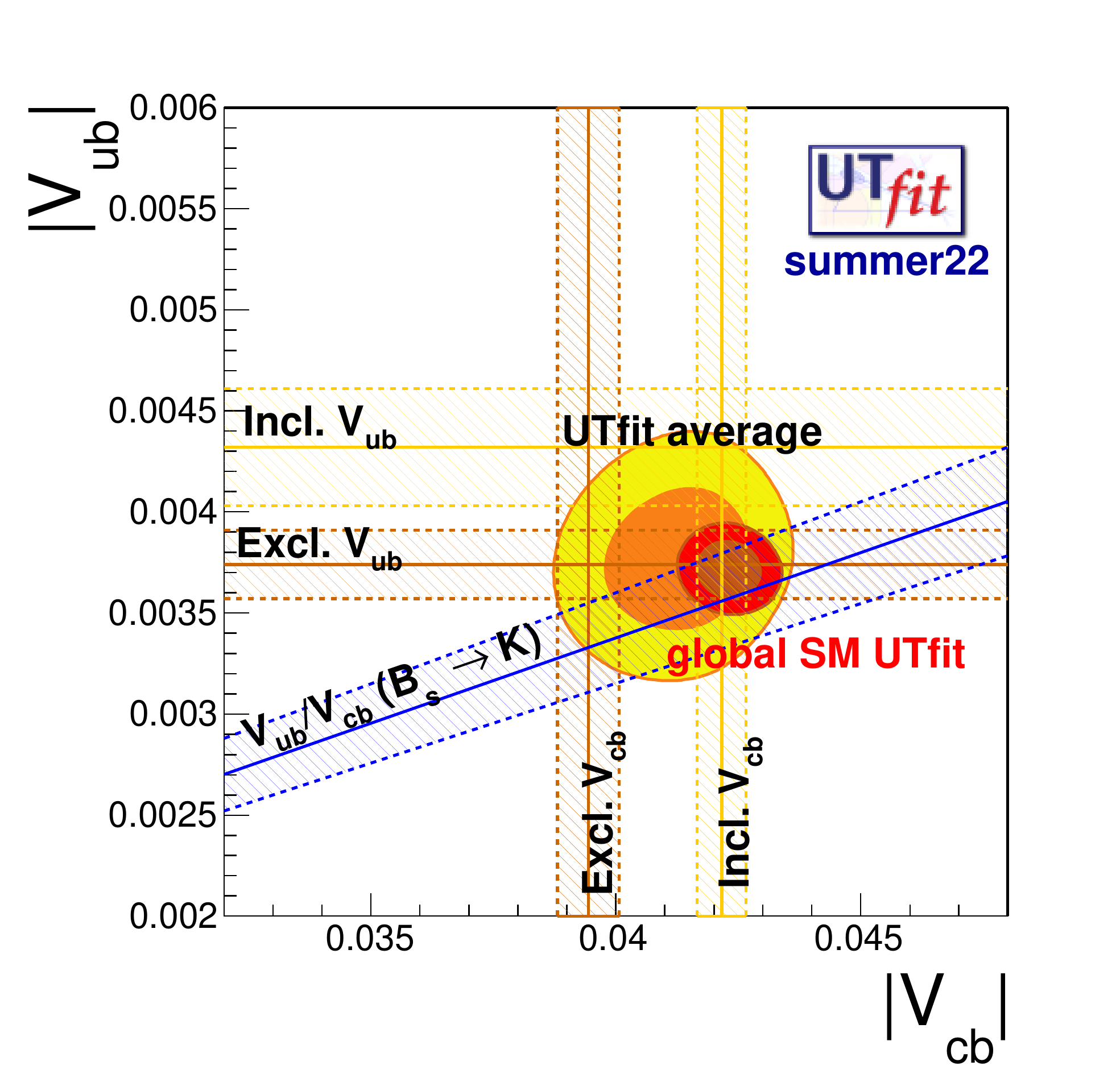}
\caption{Status of the global determination of $\textrm{V}_{\textrm{ub}}$ and $\textrm{V}_{\textrm{cb}}$. 
Reproduced from UTFit\cite{UTfit:2022hsi}.}
\label{fig:vubvcbutfit}
\end{figure}

There is a longstanding tension between the inclusive and exclusive determinations for both $|\textrm{V}_{\textrm{ub}}|$ 
and $|\textrm{V}_{\textrm{cb}}|$. The latest UTFit analysis\cite{UTfit:2022hsi} summarised in Figure~\ref{fig:vubvcbutfit} 
shows that this tension is currently greater than three standard deviations for $\textrm{V}_{\textrm{cb}}$ and $2-3\sigma$ depending on the 
precise inputs used for $\textrm{V}_{\textrm{ub}}$. The tension between inclusive and exclusive determinations is 
generally taken as a sign of imperfectly understood experimental (in the inclusive) or theoretical (in the exclusive) 
effects, rather than physics beyond the SM. It is interesting to note that recent Belle~II exclusive 
analyses\cite{Belle-II:2022imn,Belle-II:2022ffa} are in agreement with other exclusive measurements. This 
further hints that any imperfectly understood experimental effects are linked to methodology, rather than a problem 
with the BaBar or Belle detectors.

A ``third way'' towards $\textrm{V}_{\textrm{ub}}$ is offered by the purely leptonic decays 
$B^+ \to (\tau,\mu)^+ \nu_{(\tau,\mu)}$ which have much smaller theoretical uncertainties than the semileptonic 
$B^0 \to \pi^- l^+ \nu_l$ processes. Since the former proceed through an annihiliation diagram they are unfortunately 
much less statistically sensitive\cite{Workman:2022ynf,HFLAV:2022pwe}: at present around $\sim\! 9\%$ relative compared to around $\sim\! 5.5\%$ 
for the semileptonic decays. Belle~II however expects\cite{Belle-II:2018jsg} to measure $\textrm{V}_{\textrm{ub}}$ 
from each these processes with a relative uncertainty of around $2.5\%$ with its full dataset, so they will play 
an important role in the legacy understanding of the CKM unitarity triangle. The corresponding $\textrm{V}_{\textrm{cb}}$ 
processes would only be accessible at a high-luminosity $e^+e^-$ collider operating at the $Z$ pole such as FCC-ee.

\section{Searching for virtual traces of beyond SM particles}

As we saw with CKM metrology, particles outside the Standard Model can have virtual interactions with their
SM counterparts, modifying the properties of SM particles and in particular the frequency with which they
decay to certain final states. In the case of CKM metrology there are no Standard Model predictions for the
individual sides and angles, so the sensitivity to beyond SM physics comes from testing the overall consistency of
the measurements: when comparing loop-level and tree-level processes, or when testing the unitarity of the
CKM triangle. There are however rare or forbidden processes in the SM for which SM predictions do 
exist\footnote{The measured CKM triangle parameters are, of course, an input to these predictions and in
some cases their leading source of uncertainty.} and which can therefore serve as effective null tests of 
the Standard Model. Within quark flavour physics such processes play an analogous role to direct BSM 
searches in general-purpose collider experiments.

\begin{figure}[t]
\centering
\includegraphics[width=0.9\linewidth]{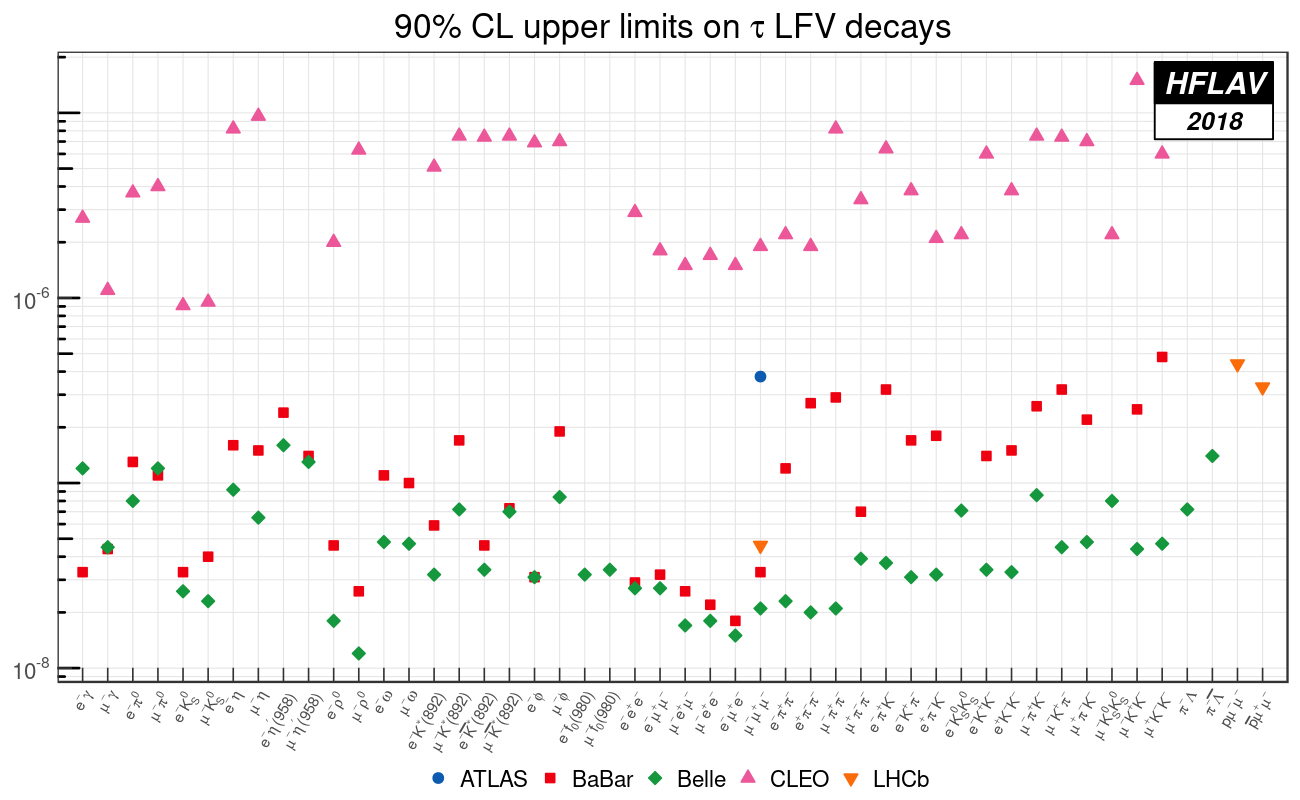}
\caption{Status of searches for lepton flavour, lepton number, and baryon number violation (as well as combinations of
these) in $\tau$ decays. Reproduced from HFLAV\cite{HFLAV:2022pwe}.}
\label{fig:lfvtau}
\end{figure}

Perhaps the most straightforward category of such measurements are searches for baryon number, lepton
number, or lepton flavour violating processes. Each of these quantum numbers is conserved in the SM while
some or all of them are generically broken in beyond SM models\footnote{Neutrino oscillations violate the
conservation of lepton flavour. However, while it is a matter of semantics whether one calls neutrino
oscillations a beyond SM process, it isn't a matter of semantics that neutrino oscillations are 
around 40 orders of magnitude too weak to induce observable lepton flavour violation in other SM processes.} 
so that any observation of these processes would be an unambiguous sign of beyond SM physics. The experimental 
approach here has been to essentially measure everything which can be measured, whether at general-purpose 
quark flavour experiments or experiments dedicated to individual processes such as $\mu\to e\gamma$ or 
$\mu\to 3e$. From an experimental point of view, this is justified since these processes are impossible in the SM
and so there is no need to care about how well the hadronic structure of any decay is understood when searching 
for them. From a phenomenological point of view, if lepton flavour violation is observed in one process then 
observing (or not observing!) lepton flavour violation in other processes is vital in order to characterize the 
flavour structure of the beyond SM physics. 
As a consequence any decay of strange, charm, or beauty hadrons which can be written down and which violates 
one of these quantum numbers should be (and largely has been) searched for. No signal has been found, and the 
overall status is summarized in the latest HFLAV compilation\cite{HFLAV:2022pwe} of the results. Figure~\ref{fig:lfvtau} 
illustrates the status of searches in $\tau$ decays, which are particularly popular since many beyond SM 
scenarios couple preferentially to the third generation of quarks and leptons.

\begin{figure}[t]
\centering
\includegraphics[width=0.9\linewidth]{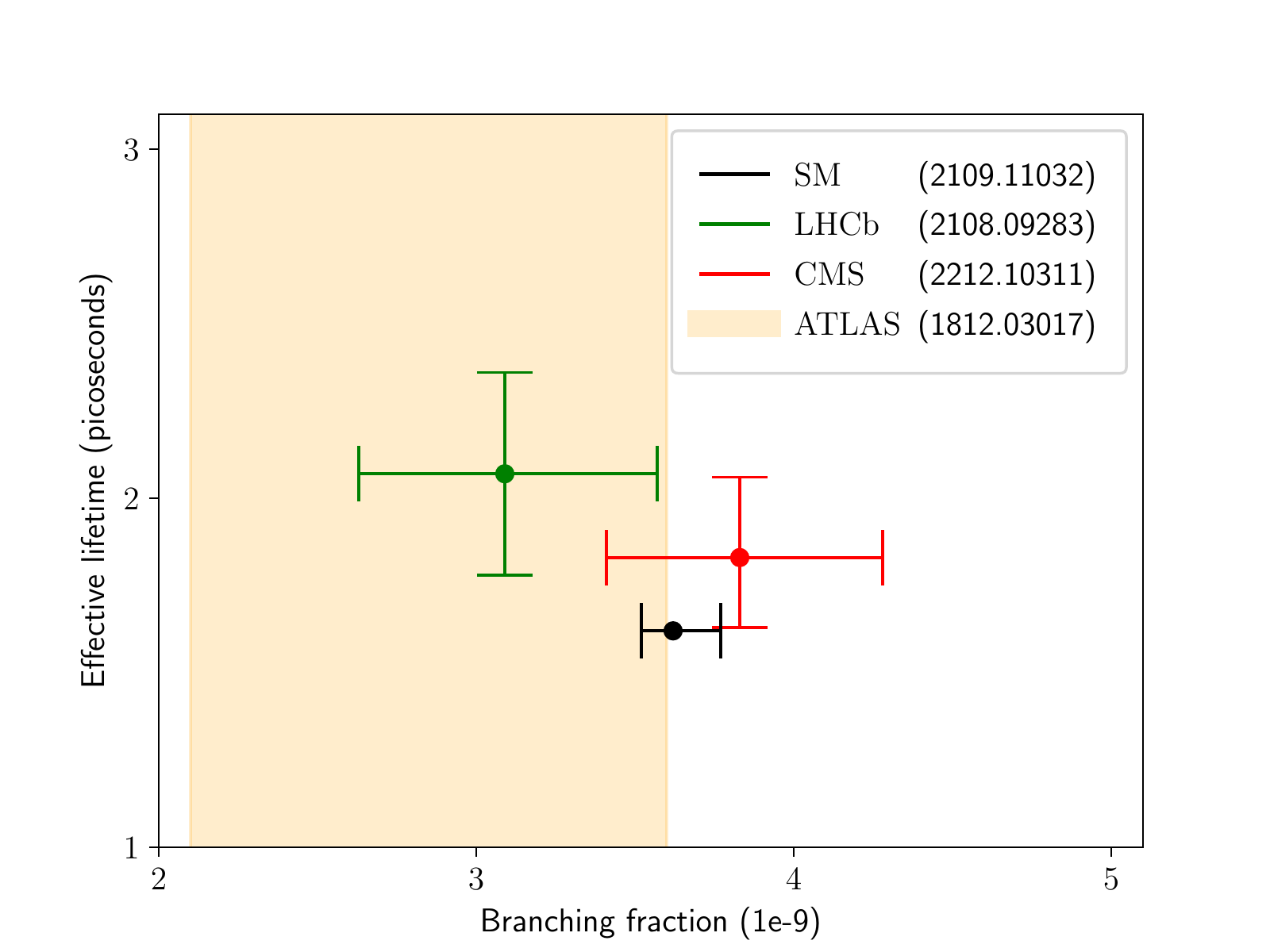}
\caption{Current measurements\cite{ATLAS:2018cur,LHCb:2021vsc,LHCb:2021awg,arxiv.2212.10311} of the branching fraction and 
effective lifetime of the $B^0_{s}\to \mu^+\mu^-$ 
decay compared to SM predictions\cite{Buras:2021nns}. Because the SM uncertainty on the effective lifetime is
almost two orders of magnitude smaller than the experimental uncertainties ($0.007$~ps) it is not plotted here for clarity. 
The experimental measurements of the branching ratio calculate their efficiency using simulation which assumes
a certain effective lifetime for the $B^0_{s}\to \mu^+\mu^-$ decay, which correlates the measurement of the
branching fraction and effective lifetime\cite{Dettori:2018bwt}. This subtle effect has a small impact given 
the current experimental uncertainties and is therefore ignored to simplify the presentation.
As ATLAS have not measured the effective lifetime their measurement is plotted as a shaded band.}
\label{fig:b2mumu}
\end{figure}

The next simplest category of null tests are (nearly) leptonic decays of Standard Model mesons which are
dominated by short-distance physics and do not contain any complex hadronic structure. These decays are 
particularly rare in the Standard Model due to GIM\cite{Glashow:1970gm} and helicity suppression, while 
they can be generically enhanced (or further suppressed) in beyond SM scenarios. In addition the rates of 
these processes can be precisely predicted in the SM, with the biggest uncertainties typically arising 
from the finite knowledge of CKM matrix elements rather than any intrinsic uncertainties on the decay 
process itself. The premier and experimentally most accessible examples of this type of process 
are\cite{Buras_2012,Buras:2021nns} $B^0_{(s)}\to \mu^+\mu^-$ and $K^+\to \pi^+\nu\bar{\nu}$. 
The $B^0_{s}\to \mu^+\mu^-$ decay has been observed by LHCb and CMS, and measurements of both its 
branching fraction and lifetime agree well with Standard Model predictions as shown in Figure~\ref{fig:b2mumu}. 
The $B^0\to \mu^+\mu^-$ decay has not yet been observed, and the global picture is one of a mild deficit with 
respect to the Standard Model. It will be interesting to see if the picture changes after Run~3 of the LHC, 
where the ATLAS, CMS, and LHCb experiments should be able to find evidence for the $B^0$ decay at its Standard 
Model value. NA62 has recently reported first evidence\cite{NA62:2021zjw} for the $K^+\to \pi^+\nu\bar{\nu}$ decay, 
with a branching fraction in good agreement with the Standard Model. The neutral kaon counterpart $K^0\to \pi^0\nu\bar{\nu}$ 
is much more challenging to access experimentally, often described as a ``nothing in, nothing out'' experiment, 
and proof-of-concept searches\cite{KOTO:2020prk} remain far from the sensitivity required to observe the Standard 
Model value. 

In leptonic beauty meson decays the corresponding electron mode is helicity suppressed so as to be unobservable 
at the SM value, and therefore acts as a pure null test of the SM. The corresponding tauonic decay is enhanced 
in the SM and even more enhanced in many beyond SM models, however the reconstruction of two tau leptons is 
experimentally challenging and leads to far greater background contamination. Searches for 
$B^0_{(s)}\to e^+e^-$\cite{Aaij:2712470} and $B^0_{(s)}\to \tau^+\tau^-$\cite{Aaij:2255070} decays 
have so far observed no signals. Meanwhile the corresponding
weak charm decays, for example $D^0\to\mu^+\mu^-$ are dominated by long-distance effects and therefore no precise
SM predictions exist. That has not stopped experimentalists from searching\cite{LHCb-PAPER-2022-029}, so far 
unsuccessfully, for this decay using the abundant $c\bar{c}$ production at the 
LHC and the narrow $D^{*0}\to D^0\pi$ decay chain to suppress combinatorial backgrounds. Recent searches for leptonic decays
of $D^{*0}$ mesons, which are not suppressed by helicity but rather by the fact that the $D^{*0}$ can decay
through the strong and electromagnetic interactions, have set\cite{LHCb:2023fuw} stringent limits but failed to find
any signals. Analogous decays involving more than two leptons in the final state can get bigger enhancements
in some beyond SM scenarios, but searches for these\cite{LHCb:2021iwr} have so far come up emptyhanded as well.

\begin{figure}[t]
\centering
\includegraphics[width=0.7\linewidth]{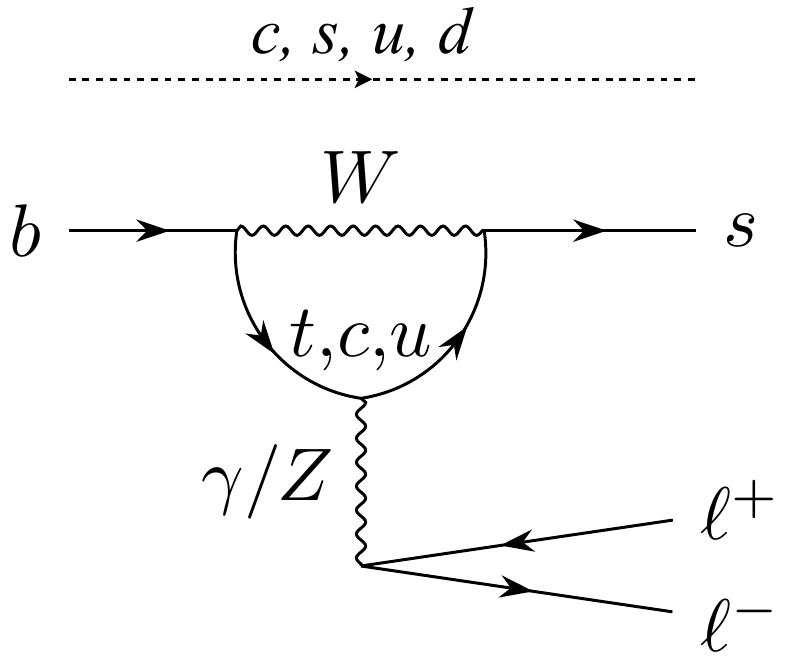}
\caption{Feynman diagram of the leptonic $B^0_{s}\to \mu^+\mu^-$ and the semileptonic $B^0\to K^{*0} \mu^+\mu^-$ decays,
where the role of the additional spectator quark in semileptonic decays is indicated using a dashed arrow.}
\label{fig:b2sllfeynman}
\end{figure}

\begin{figure}[t]
\centering
\includegraphics[width=0.9\linewidth]{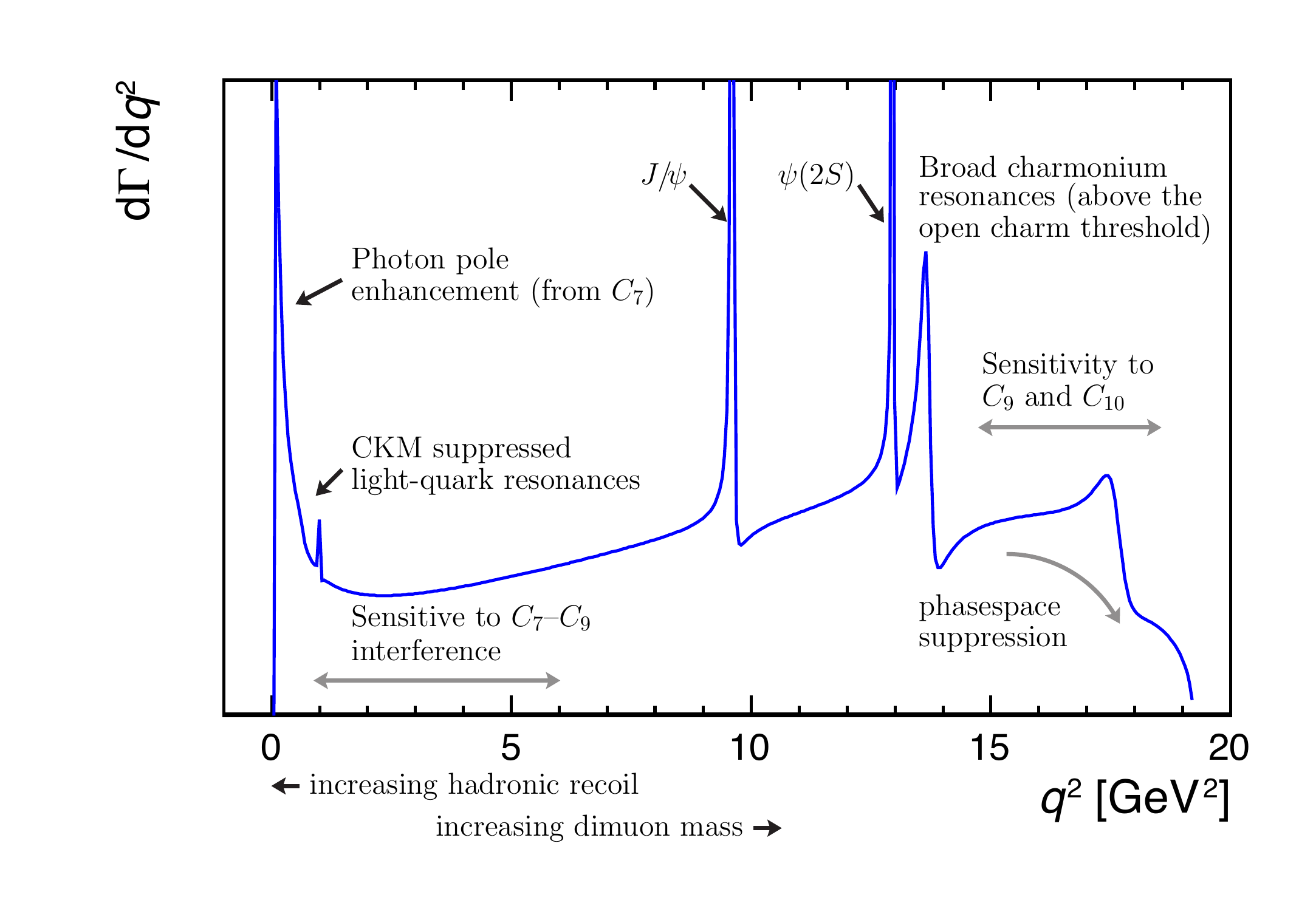}
\caption{Cartoon illustrating the dimuon mass squared, $q^2$, dependence of the differential decay rate of 
$B \to K^{*}\ell^+\ell^-$ decays. The different contributions to the decay rate are also illustrated. For 
$B \to K\ell^+ \ell^-$ decays there is no photon pole enhancement due to angular momentum conservation. 
Caption and illustration reproduced from Blake Lanfranchi and Straub\cite{Blake_2017} with permission 
of the authors.}
\label{fig:b2sllwilsoncoeffs}
\end{figure}

Semileptonic loop-level decays are both less rare, with branching fractions of up to $10^{-6}$, and more 
phenomenologically complex than their leptonic counterparts. As an example, consider the $b \to s l^+ l^-$
process which is the most extensively studied of its kind. The Standard Model Feynman diagrams for the
leptonic $B^0_{s}\to \mu^+\mu^-$ and the semileptonic $B^0\to K^{*0} \mu^+\mu^-$ decays are compared
in Figure~\ref{fig:b2sllfeynman}. As we can see these diagrams differ only in the fate of the spectator quark, which
hadronizes in the semileptonic case and is indirectly annihilated in the leptonic case. The additional
final state hadron nevertheless leads to much richer decay dynamics depending on the invariant mass of the
dilepton pair, commonly referred to as the $q^2$ of the decay. 
This dependence is summarized in Figure~\ref{fig:b2sllwilsoncoeffs}.
The decay dynamics also means that the differential decay rate is, depending on the quantum numbers of the
final state hadron, a function of the angles between the final-state particles in the $b$-hadron rest frame.
The Standard Model prediction for the differential decay rate suffers from significant uncertainties 
due to poorly-known hadronic form factors. The angular dependence however makes it possible to construct a 
number of observables\cite{Descotes-Genon:2013vna} where the leading-order form-factor uncertainties cancel. 
Furthermore, the overall set of angular and $q^2$ observables can be interpreted in terms of the Wilson 
coefficients, which encode the short-distance physics, and compared to Standard Model predictions. These 
so-called ``global fits'' have been very popular in recent 
years\cite{Alguero:2023jeh,Ciuchini:2022wbq,Gubernari:2022hxn,Mahmoudi:2022lmp,Altmannshofer:2021qrr,Descotes-Genon:2020buf} 
and test not only specific observables but the overall coherence of 
the measured observables within one $b \to s l^+ l^-$ process as well as between different $b \to s l^+ l^-$
processes. They can also be used\cite{Greljo:2022jac} to interpret the $b \to s l^+ l^-$ observables
within the SMEFT\cite{Brivio:2017vri} and derive corresponding bounds on high-energy processes which can be
confronted with direct searches at the LHC and future colliders. Since all $b \to s l^+ l^-$ processes 
should be governed by the same short-distance dynamics and hence the same Wilson coefficients whether in the 
Standard Model or its extensions, the overall fit quality is also a good indicator
of potential incoherences in the experimental measurements -- similarly to global fits of the CKM triangle.

From an experimental point of view, the different $q^2$ regions are easy to separate in the case of muonic
decays but more complex for electrons due their much worse momentum resolution and Bremsstrahlung energy loss.
Tauonic decays are even more complex due to the presence of multiple neutrinos in the final state and the
associated energy losses, and neither Belle~2 nor LHCb will have the sensitivity to observe them at the
Standard Model value. For these reasons the bulk of experimental results have used $b \to s \mu^+ \mu^-$
decays, with experiments only just beginning to be sensitive to the CKM-suppressed but otherwise 
analogous $b \to d \mu^+ \mu^-$ processes. The exception are tests of electron-muon universality in the
integrated branching fractions of $b \to s \mu^+ \mu^-$ and $b \to s e^+ e^-$ decays, which have been studied
extensively in recent years.

While $b \to s \mu^+ \mu^-$ processes were observed by BaBar and Belle, it was not until
LHCb began datataking in earnest in 2011 that the full range of $q^2$ dependent observables could be precisely
measured. It quickly became apparent that their branching fractions were below Standard Model predictions
across most of the $q^2$ range, and this ``muon deficit'' has so far been observed for all $b \to s \mu^+ \mu^-$ 
processes studied by LHCb. Measurements from other collaborations, though significantly less precise, also
generally agree with this trend. More intriguingly, deviations from SM predictions were also spotted in angular 
observables, with the biggest deviation seen in the so-called $P^{'}_{5}$ observable in $B^0\to K^{*0} \mu^+\mu^-$,
for which the leading-order form-factor uncertainties cancel. As with the muon deficit, coherent deviations in 
angular observables have been seen across all studied $b \to s \mu^+ \mu^-$ processes. 

\begin{figure}[t]
\centering
\includegraphics[width=0.7\linewidth]{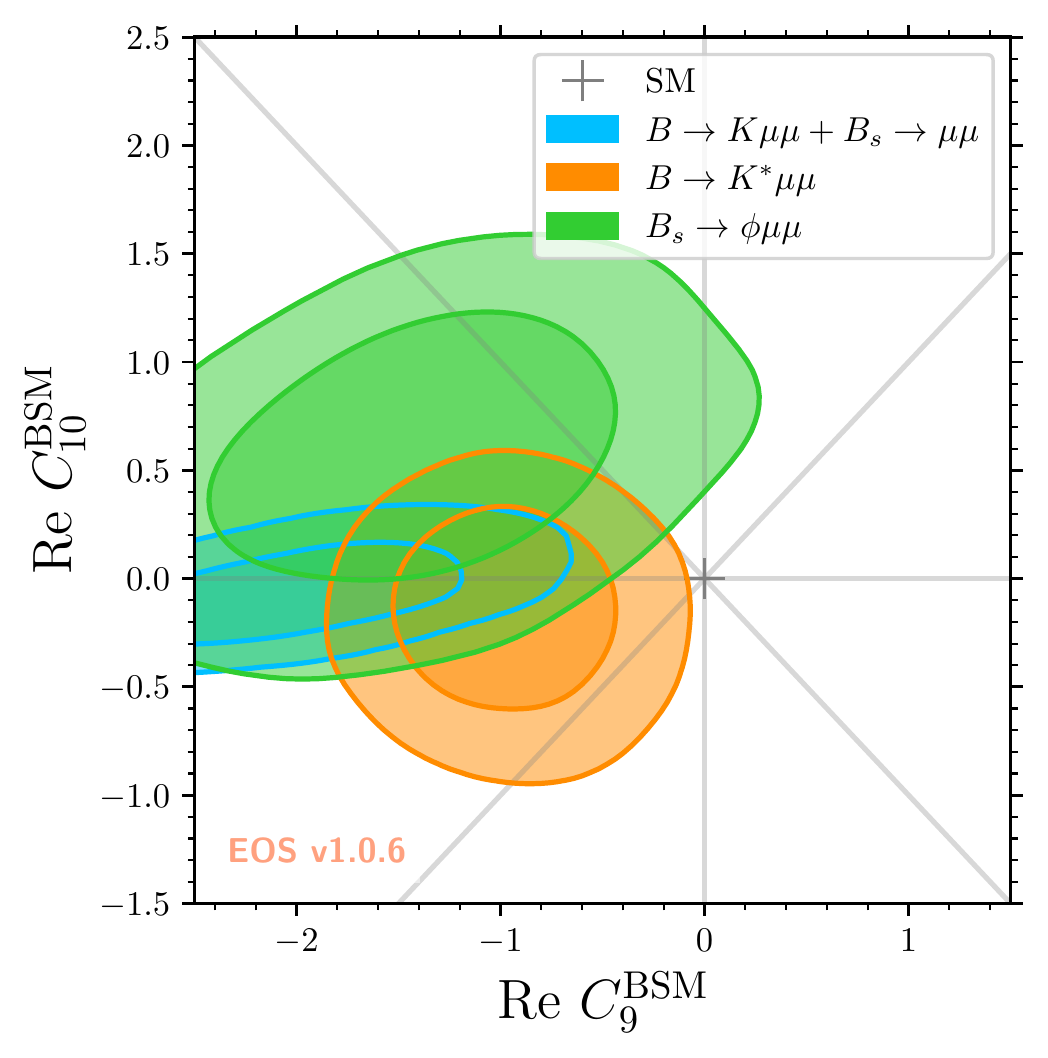}
\caption{Two dimensional fit of experimental $b\to s\ell\ell$ measurements to the muonic $C_9$ and $C_{10}$ 
Wilson coefficients. The expected SM value is $(0,0)$,
while contours arising from different experimental measurements of $b \to s \mu^+ \mu^-$ processes are
described in the legend. This plot was updated from the one in Gubernari et al\cite{Gubernari_2022}
by M\'{e}ril Reboud for this review.}
\label{fig:b2sllglobalfits}
\end{figure}

As mentioned earlier it is popular to represent the experimental status through global fits to the 
muonic $C_9$ and $C_{10}$ Wilson coefficients, with one such recent fit shown in Figure~\ref{fig:b2sllglobalfits}. 
\footnote{Constraints on the $C_{10}'$ Wilson coefficient are used instead of $C_{10}$ when 
testing models which predict new right-handed currents, but the picture of agreement with respect to the 
Standard Model remains basically the same.} These global fits typically 
include not only semileptonic $b \to s \mu^+ \mu^-$ processes but also other measurements sensitive to the 
same Wilson coefficients, most notably the branching fraction of $B^0_{s}\to \mu^+\mu^-$ which constrains $C_{10}$.
As we can see, $C_{10}$ is presently compatible with Standard Model predictions, while $C_9$ can deviate 
significantly from the Standard Model. The precise deviation depends on the assumptions made in the global 
fits about the residual hadronic uncertainties in the semileptonic processes, but is frequently found to be 
greater than the canonical $5\sigma$ discovery threshold. So why has the field not declared a discovery of 
beyond SM contributions to these processes?

\begin{figure}[t]
\centering
\includegraphics[width=0.95\linewidth]{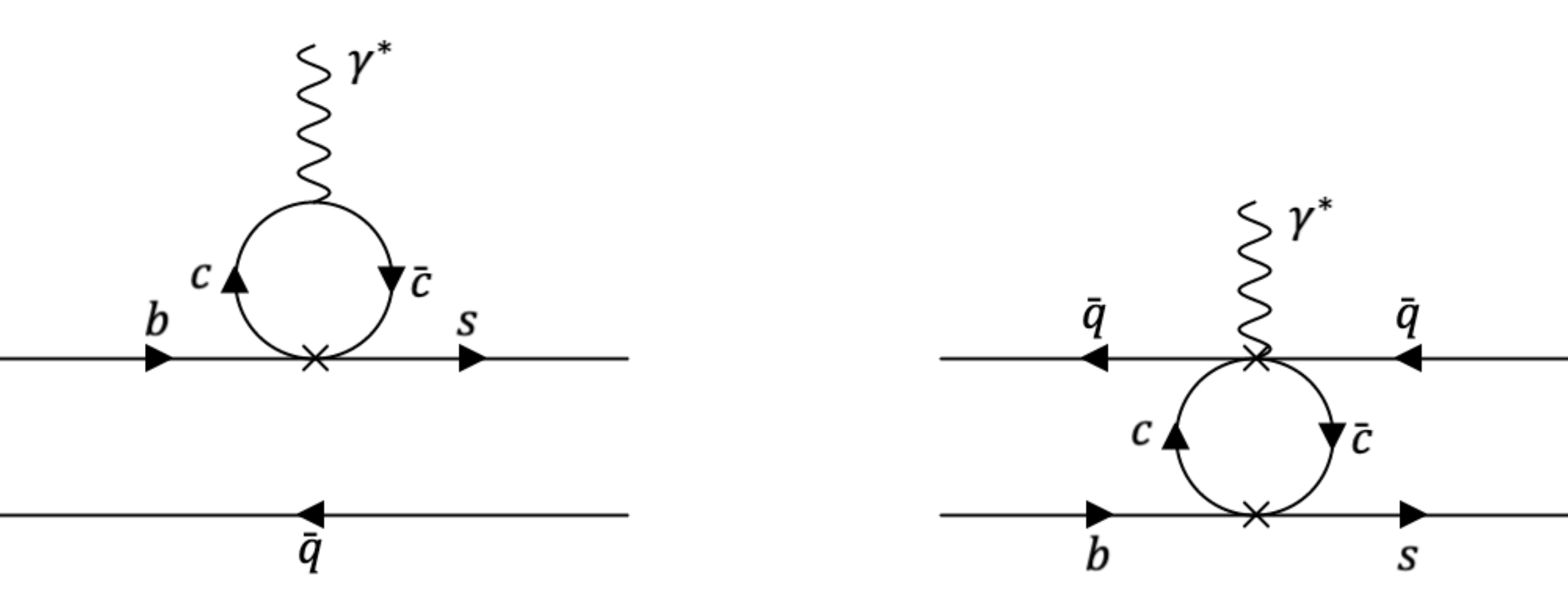}
\caption{Example Feynman diagrams for charm loop processes in which (left) the $c-\bar{c}$ loop directly produces a (virtual) 
photon and (right) in which the virtual photon is produced after a rescattering involving intermediate hadronic states. 
In both cases the result is a long-distance contribution to the overall decay rate, which can interfere with the 
short-distance electroweak process that is of interest when testing the Standard Model.}
\label{fig:charmloops}
\end{figure}

The first signs of significant deviations in angular $b \to s \mu^+ \mu^-$ observables sparked an intense
theoretical\cite{Khodjamirian:2012rm,Jager:2012uw,Fajfer:2012nr,Lyon:2014hpa,Ciuchini:2015qxb,Bharucha:2015bzk,Capdevila:2017ert,Jager:2017gal,Khodjamirian:2017fxg} and 
experimental\cite{LHCb:2013ywr,Aaij:2016cbx,LHCb:2020puj,LHCb:2020pxc,LHCb:2020gnv,LHCb:2022vsv,LHCb:2022bkt} 
effort to understand whether poorly understood hadronic dynamics 
could be causing the observed discrepancies with respect to Standard Model predictions. Non-trivial hadronic
structures indicative of exotic $c\bar{c}$ states which can decay into a pair of muons have been observed
above the open charm kinematic threshold, and the $q^2 <1$~GeV$^2$ region is polluted by numerous light hadronic
resonances which can decay into dileptons with non-negligible branching fractions. However the region with
the biggest observed discrepancies, roughly speaking $1< q^2 <6$~GeV$^2$, lies below the open charm threshold
and does not contain any known hadronic resonances which can decay into a dimuon pair. Nevertheless, as
discussed extensively in the literature, so-called ``charm loop'' diagrams illustrated in Figure~\ref{fig:charmloops}
can contribute to the $b \to s \mu^+ \mu^-$ decay rate even in this region. Since these are long-distance 
diagrams they can mimic the effect of beyond SM physics, and some 
parametrisations\cite{Ciuchini:1997hb,Ciuchini:2022wbq} of the charm
loops can fully explain the observed deviations from the SM. The precise size of these charm loop
contributions, and the best way to parameterize them, are however currently open questions. 

One way forward is to attempt to measure the size of charm loop contributions directly from data. For example
LHCb measured the interference between short-distance and charm-loop contributions in $B^+\to K^+ \mu^+\mu^-$
decays, finding it to be minimal. LHCb is also currently performing analyses of $B^0\to K^{*0} \mu^+\mu^-$ decays 
using two different proposed\cite{Chrzaszcz_2019,blake2017empirical} charm loop models. Charm loops can generically lead to $CP$ violating
effects, while no $CP$ violation has been obsered in $b \to s \mu^+ \mu^-$ to date. These results and their
interpretation depend on the theory model used, and their interpretation is also subject to an understandable
bias in favour of the Standard Model, particularly for measurements affected by even hypothetical hadronic
effects. That is to say that if the experimental picture shows hints of large charm loop effects, this result 
will likely be taken at face value. On the other hand even if the experimental picture remains coherent 
and continues to point\cite{Gubernari_2022} to charm loop effects which are too small to explain the observed 
deviations, it is not clear whether that will lead to a consensus that beyond SM physics has been observed.

None of the caveats around charm loops affect the interpretation of lepton universality tests in semileptonic 
$b \to s l^+ l^-$ processes. The ratio of electron and muon rates can be theoretically predicted with
percent level precision\cite{Bordone:2016gaq} within the Standard Model, and even this uncertainty comes from the modelling 
of QED effects in the electron modes rather than from any effects related to the hadronic part of the decays. 
If beyond SM Wilson coefficients are present the hadronic unceratinties no longer perfectly cancel, however this
only complicates the interpretation of any putative beyond SM effects. There is universal
consensus that any observation of lepton universality breaking in these decays would be an unambiguous sign
of physics beyond the Standard Model. In 2014 LHCb reported a measurement\cite{LHCb:2014vgu} of electron-muon universality
in the $1< q^2 <6$~GeV$^2$ region of $B^+ \to K^+ l^+ l^-$ decays which deviated from SM predictions by 
around $25\%$ or $2.5\sigma$. LHCb followed this up in 2017 with the analogous measurement\cite{LHCb:2017avl} in 
$B^{*0} \to K^{*0} l^+ l^-$ processes which showed coherent deviations of a similar size. In both cases 
electrons were more abundant than muons, or in other words, while muon branching fractions showed a deficit 
with respect to SM predictions, the electron branching fractions agreed well with them. And in 2019 LHCb published
the first test\cite{LHCb:2019efc} of electron-muon universality in baryonic $b \to s l^+ l^-$ decays which was compatible
with both the Standard Model and with the deviations seen in the previous analyses. 

A puzzling aspect of the 2017 $B^{*0} \to K^{*0} l^+ l^-$ analysis was a $\approx\! 2\sigma$ deviation from 
lepton universality in the $0.045 < q^2 < 1.1$~GeV$^2$ region, where the decay rate is dominated by a virtual 
photon decaying into the dilepton pair\footnote{This region is commonly known as the ``photon pole'' in the 
literature and only occurs for those decays where it is allowed by spin-parity considerations.} which must 
respect lepton universality not only in the Standard Model but in any beyond SM scenario which does not involve 
the on-shell production of beyond SM states. However when the 2021 analysis\cite{LHCb:2021trn} of electron-muon 
universality in $B^+ \to K^+ l^+ l^-$ decays using LHCb's full dataset showed a $3.1\sigma$ deviation of around
$15\%$ from the Standard Model, and a subsequent 2022 analysis\cite{LHCb:2021lvy} of $B^+ \to K^{*+} l^+ l^-$ and 
$B^0 \to K^0_S l^+ l^-$ decays also showed coherent deviations, the community struggled to contain its 
excitement.

It is important to underline that, unlike the angular observables, there was nobody claiming that these 
results could be caused by poorly understood hadronic effects, or by any other kinds of effects linked to 
the phenomenological interpretation of these processes. In fact, the one sceptical paper written during this 
period\cite{Robinson:2021cws} raised concerns about the experimental treatment of the electron modes rather than 
about the interpretation of the results. While it was uncomfortable that no experiment other than LHCb had the 
sensitivity to truly cross-check these results, measurements performed by Belle seemed to find similar 
deviations\cite{Belle:2016fev,Belle:2019oag} as the ones observed by LHCb.  

\begin{figure}[t]
\centering
\includegraphics[width=0.8\linewidth]{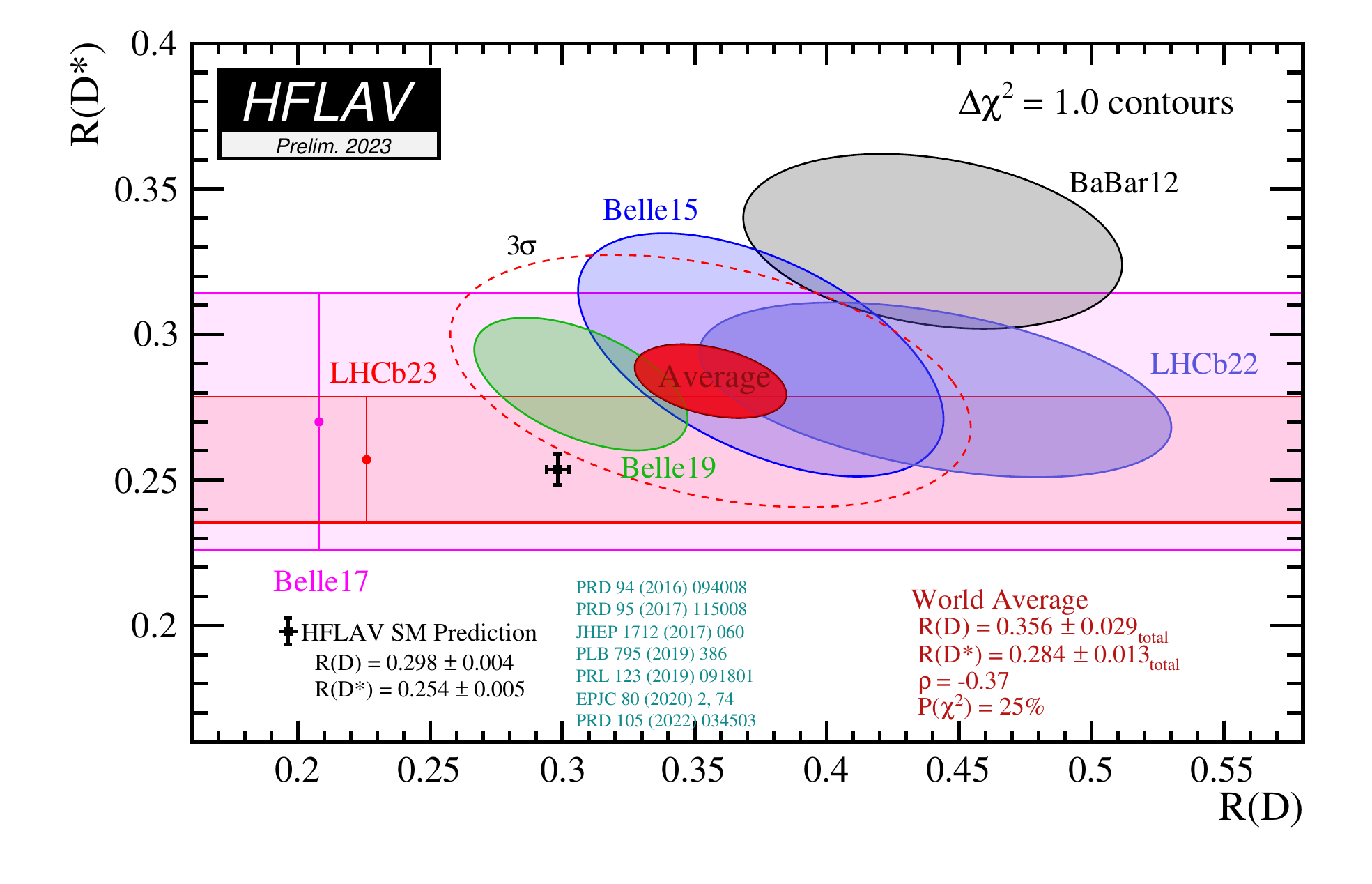}
\caption{Current experimental constraints in the $\textrm{R}(\textrm{D})-\textrm{R}(\textrm{D}^*)$ plane compared with the 
Standard Model prediction. Reproduced from HFLAV\cite{HFLAV:2022pwe}.}
\label{fig:rdrdst2023}
\end{figure}

Before addressing the fate of lepton universality in $b \to s l^+ l^-$ processes, it is worth taking a detour
to consider the tree-level flavour-changing $b\to c l^- \nu$ transitions. We
have already discussed these decays in the context of CKM measurements, but they can also be used to perform
precise tests of lepton universality. While electron-muon universality tests in $b\to c l^- \nu$ transitions
agree with the Standard Model prediction at the percent level\cite{Belle-II:2023qyd} and are generally limited by experimental
systematics, the same is not the case for universality tests between the tau and the two lighter leptons.
Ever since BaBar's original evidence\cite{BaBar:2012obs} for an excess of $\bar{B}\to D^{(*)}\tau^-\bar{\nu}$ decays, 
a number of measurements by BaBar\cite{BaBar:2013mob}, Belle\cite{Belle:2015qfa,Belle:2016ure,Belle:2019gij}, 
and LHCb\cite{LHCb:2015gmp,LHCb:2017smo,LHCb:2017rln,LHCb:2023cjr} have shown signs of a coherent 
deviation from lepton universality in these processes, as illustrated in the latest HFLAV world average in
Figure~\ref{fig:rdrdst2023}. The labels $\textrm{R}(\textrm{D})$ and $\textrm{R}(\textrm{D}^*)$ refer to
lepton universality tests of the beauty meson decaying into the ground-state charm meson or the lowest
excited charm meson state. Since the $D^*$ can decay into the ground state meson it is possible to measure
the $D^*$ channel separately, but the ground-state channel can only be measured together with the excited
one. Higher charm excitations also contribute to these decays but are generally treated as backgrounds.

From an experimental point of view, BaBar and Belle (as well as Belle~2) are intrinsically better suited to 
the study of tree-level semileptonic processes because of their hermetic detectors and because their events
contain a $b\bar{b}$ meson pair and nothing else. This allows the missing energy associated with the neutrino
emitted in the $b\to c l^- \nu$ process to be constrained by measuring the energy of the ``other'' $B$ meson 
in the event and subtracting it from the known beam energy. There are two principal methods to measure the
energy of the other $B$. The first, known as the semileptonic tag, uses the charged lepton emitted in a
$b\to c l^- \nu$ decay as a proxy for the $B$ meson energy. The second, known as the hadronic tag, fully
reconstructs a collection of $O(1000)$ $B$ meson decays to determine the $B$ meson energy. The semileptonic
tag method generally has better efficiency, while the hadronic has better purity. The performance for electrons 
and muons is also sufficiently similar that both can be used in the measurement, increasing statistics and 
allowing an electron-muon universality test to be performed as a byproduct of the tau measurement. The only
published BaBar result uses hadronic tagging, while Belle has published both a hadronic and 
a semileptonic tagged result, as well as a measurement which reconstructs the signal tau 
lepton through its hadronic decays $\tau\to\pi\nu_\tau$ and $\tau\to\rho\nu_\tau$.

At LHCb the missing neutrino energy can be inferred from the difference between the reconstructed $B$ meson
momentum and displacement vectors\cite{Ciezarek:2016lqu} while the tau lepton can be reconstructed in the 
$\tau\to\mu\nu_\tau\bar{\nu_\mu}$ and $\tau\to 3\pi\nu_\tau$ final states. The former has the advantage of
a direct normalization to the $b\to c \mu^- \nu_\mu$ decay, while the latter has the advantage that the
tau decay vertex formed by the three charged pions, as well as the known Dalitz structure of the tau decay,
help to reject backgrounds. Since the detector performance and backgrounds are substantially different in these two
cases, their compatibility reinforces the experimental robustness of the results similarly to the use of
hadronic or leptonic opposite-side tags to reconstruct these decays at BaBar, Belle, and Belle~2. 
LHCb has published measurements in the muonic and hadronic final states, as well as a 
measurement\cite{LHCb:2022piu} of the analogous baryonic process $\Lambda_b\to\Lambda_c l^- \nu$ in 
the hadronic final state.

\FloatBarrier

\begin{figure}[ht]
\centering
\includegraphics[width=0.95\linewidth]{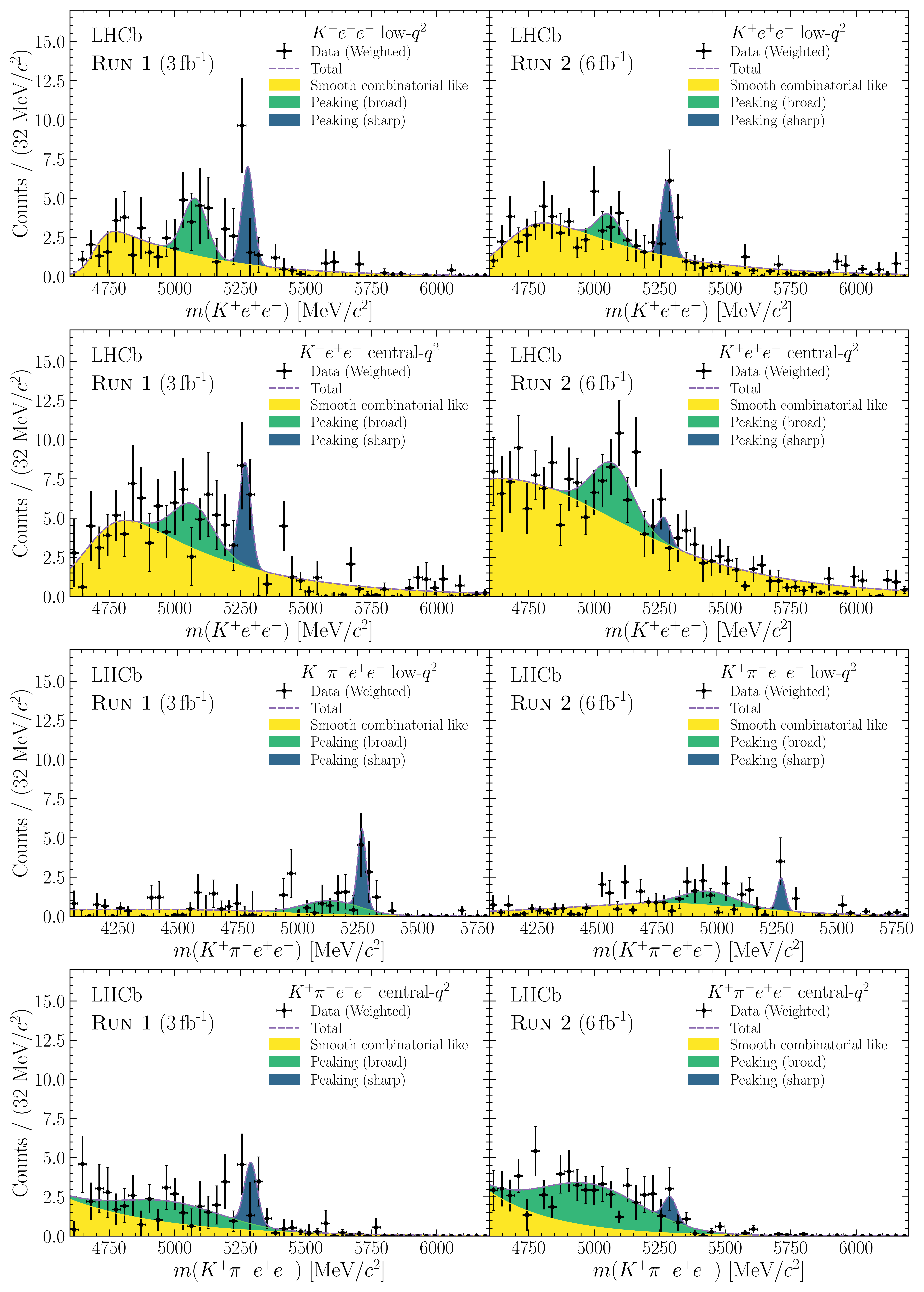}
\caption{Template shapes for misidentified backgrounds obtained from data. The shapes for Run 1 are given on the 
left, the shapes for Run 2 are given on the right. From top to bottom, the shapes for $R_K$ in low $q^2$, $R_K$ in 
central $q^2$, $R_{K^{*0}}$ in low $q^2$, and $R_{K^{*0}}$ in central $q^2$ regions are given. Caption and 
illustration reproduced from Ref.~\cite{arxiv.2212.09153}.}
\label{fig:misidRX}
\end{figure}

\FloatBarrier

At present the experimental measurements are statistically limited, and most of LHCb's dataset has yet to
be analysed. The current world average is around $12-20\%$ from the SM value depending on the charm meson 
in the final state. We can expect the muonic LHCb measurement to gain roughly a factor two in sensitivity, and the hadronic
a factor of $\sqrt{2}$, even before data from the upgraded LHCb detector is considered.
In addition the $B^0_s$ analogues of these LUV tests, $\textrm{R}(\textrm{D}_s)$ and $\textrm{R}(\textrm{D}^*_s)$, 
as well as the charged charm analogue $\textrm{R}(\textrm{D}^+)$, can also be measured by LHCb and would 
add complementary information. All of these LUV tests will be performed with both the muonic and hadronic
decays of the $\tau$. If the charged-current anomaly persists at roughly the current central values, 
Belle~2 and LHCb should both be able to make single-experiment observations of LUV over the next decade, which 
would be crucial to give confidence in the experimental robustness of the results. In this context the development 
of cross-experiment tools for coherently simulating the impact of beyond SM physics on the kinematics and geometry 
of these decays, such as HAMMER\cite{Bernlochner:2020tfi}, is also of great importance. 

We can now return to the $b \to s l^+ l^-$ lepton universality tests. 
It is not trivial to explain the $b \to s l^+ l^-$ and $b\to c l^- \nu$ anomalies in the
same theoretical framework, since one set of anomalies requires a significant modification to light-lepton
decay rates in loop processes while the other requires a significant modification to tau-lepton decay
rates in tree-level processes. Nevertheless combined intepretations were gradually proposed\cite{Buttazzo:2017ixm}, 
with the common feature that they predicted sizeable effects in loop-level processes involving tau leptons as well as
in the $\tau^+\tau^-$ mass spectrum produced at the LHC. 

In December 2022 LHCb 
published\cite{arxiv.2212.09153} a combined test of electron-muon universality in $B^+ \to K^+ l^+ l^-$ and
$B^{*0} \to K^{*0} l^+ l^-$ decays using its full available dataset. Both channels were studied in two $q^2$ 
regions: $1< q^2 <6$~GeV$^2$ and $0.1< q^2 <1.1$~GeV$^2$, leading to four total measurements of lepton 
universality. All the results were fully compatible with SM predictions, with the largest single deviation
being around $1\sigma$. Compared to previous LHCb measurements, this analysis uses 
a more accurate treatment of backgrounds in which a hadron is misidentified as an electron. First of all, 
more stringent identification criteria are applied, inherently reducing the impact of such backgrounds. In addition, 
residual misidentified backgrounds are explicitly modelled in the invariant mass fit which measures the 
electron mode signal yields. Data in a control region enriched 
in such misidentified backgrounds are extrapolated into the signal region using transfer maps computed from data 
calibration samples. 

The remaining backgrounds which enter the final fit are shown in Figure~\ref{fig:misidRX}. As can be seen they
contain peaking shapes which could not be absorbed by other fit components in the previous analyses and therefore
biased them away from the SM. This data-driven treatment of backgrounds allows all four LU measurements to remain 
statistically limited, which gives confidence that LHCb should eventually be able to test electron-muon universality 
in $b \to s l^+ l^-$ to the percent level, matching uncertainties on the SM predictions.
Belle~2 will also be able to contribute to these tests\cite{arxiv.2206.05946} albeit with reduced sensitivity, 
though its real strength in this area lies in the unique and phenomenologically complementary ability to measure 
$b \to s \nu_\ell\bar{\nu_\ell}$ processes\cite{Belle-II:2018jsg,PhysRevLett.127.181802}. From a phenomenological point
of view LHCb now observes that both electron and muon branching ratios are below SM predictions, which could be
caused by QCD uncertainties or by lepton-universal beyond SM contributions. It will be interesting to see if 
lepton universality is also found to hold in the angular analysis of these decays.

Before closing, a word about a couple of other related SM tests. The charm equivalent of $b\to s l^+ l^-$
processes, $c\to u l^+ l^-$, are more difficult to interpret because of a large number of overlapping and broad
hadronic resonances which decay into dileptons within the available phase space of the decay. In addition, for
$b\to s l^+ l^-$ processes these hadronic resonances do not dominate the decay rate even at $q^2 < 1$~GeV.
In the charm case, however, the decay rate involving intermediate hadronic resonances is around three orders 
of magnitude greater than the rate of the SM leptonic flavour-changing neutral current. There are regions of
$q^2$, in particular below around 500~MeV, where the leptonic current dominates, but they are much smaller than
in the $b\to s l^+ l^-$ case. For all these reasons, while preliminary measurements have observed $c\to u \mu^+ \mu^-$
processes in the region of the hadronic resonances\cite{LHCb:2017uns,LHCb:2018qsd}, no measurement of the equivalent 
electron process or observation of the SM leptonic current exist. The family of radiative $b \to s \gamma$
decays has been extensively studied for similar reasons as the $b \to s l^+ l^-$ decays, although they
are sensitive to different Wilson coefficients and could therefore be affected differently by beyond SM physics.
The current measurements are entirely compatible with the Standard Model. The width splitting between mass 
eigenstates of the $B^0$ meson, $\Delta\Gamma_d$ has been advocated\cite{Gershon:2010wx} as a null test of the Standard Model 
due to its extremely small SM value; current measurements show no evidence for an anomalously large value.   

\section{Future experiments and facilities}

The landscape of quark flavour physics experiments and facilities was summarized in Figure~1 of a recent Snowmass 
submission\cite{Altmannshofer:2022hfs}. As things stand, the principal general-purpose flavour factories
are LHCb and Belle~2, approved to collect around 50~fb$^{-1}$ of $pp$ and 50~ab$^{-1}$ of $ee$ collision data respectively 
by roughly the early 2030s. As discussed in the physics case for a second upgrade of LHCb\cite{LHCb:2018roe}, their datasets are highly 
complementary. The general-purpose (but limited in quark flavour reach) LHC detectors ATLAS and CMS are approved 
to collect 3~ab$^{-1}$ by the early 2040s. A detailed breakdown of observables and sensitivities is given in 
Table $10.1$ of the physics case for a second upgrade of LHCb, as well as in the yellow report on the physics
of the HL-LHC~\cite{Cerri:2018ypt} and FCC~\cite{FCC:2018byv} and I will not reproduce those here but rather make a few general 
observations. 

For what concerns the apex of the unitarity triangle, the current LHCb and Belle~2 programmes will constrain
it at the $1-2\%$ level. The data on strong phases in charm hadron decays collected by the more specialised
BESIII experiment at the $\psi(3770)$ resonance will be critical in ensuring the most precise possible 
measurement of the CKM angle $\gamma$; BESIII will also have unique capabilities in the exploration of lighter 
exotic hadrons. Lepton universality in tree- and loop-level beauty decays will similarly be probed at the 
few percent level, while $CP$ violation in the charm sector will be probed at the $10^{-4}$ level in decays 
and a few times $10^{-5}$ level in the interference of mixing and decay. For what concerns measurements
of mixing in $B^0_s$ meson decays and rare leptonic processes such as $B^0_{(s)}\to\mu^+\mu^-$, ATLAS and CMS
will also provide highly complementary sensitivity and for example enable around a 20\% measurement of the
$B^0\to\mu^+\mu^-$ branching ratio.
 
While local form factors will remain a significant source of uncertainties for many observables in (semi)-hadronic
processes, many key observables will be limited by experimental statistics rather than irreducible experimental
systematics or by theoretical uncertainties. Moreover certain key observables will remain very poorly measured,
most notably beauty hadron decays to final states involving multiple $\tau$ leptons and decays of $B_c$ mesons.
These facts motivate a further generation of quark flavour experiments. A proposed second upgrade of LHCb
aims to reach around 300~fb$^{-1}$ by the early 2040s, the same timescale as the ATLAS and CMS HL-LHC upgrades.
And the detectors proposed for the FCC-ee electron collider\footnote{Broadly similar capabilities could of course
be achieved at the CEPC version of the collider.} are planned with a quark flavour physics programme
at the $Z$ pole in mind, aiming at around $10^{12}$ $Z\to b\bar{b}$ decay pairs with $O(1)$ detection and
reconstruction efficiency. 

The FCC-ee, Belle~2, and LHCb Upgrade~2 sensitivities to key observables are compared in Chapter~7 of the 
report on physics opportunities at the FCC and as earlier I will restrict myself to a few observations. 
The apex of the CKM triangle will be independently characterised at the permille level by the two facilities, 
providing not only additional reach but also a crucial ability to cross-check each other at these unprecedented 
precisions. Since they will not run concurrently such an ability will rely on the proper preservation of LHCb 
datasets for future reanalysis in light of, or concurrently with, FCC-ee datasets. The benefits of concurrent 
analysis for CKM studies have recently been gaining attention in the case of LHCb, Belle~2, and BESIII 
and similar opportunities may present themselves with FCC-ee. Electron-muon universality in $b\to s l^+ l^-$ 
processes will be probed to the percent level, likely reaching fundamental experimental systematic limitations. 
FCC-ee however will have unique sensitivity to decays involving multiple tau leptons, inclusive and leptonic 
decays of $B_c$ mesons, and allow tests of muon-tau universality in $b\to s l^+ l^-$ decays at the few percent 
level for the first time. It will also surpass the sensitivity of Belle~2 for $b\to s\nu\bar{nu}$ observables. 
In the charm sector the second upgrade of LHCb will have its own unique reach, achieving $10^{-5}$ sensitivity 
to $CP$ violation in the interference of decay and mixing and measuring charm mixing to better than the 
$10^{-5}$ level. This raises the tantalising prospect of observing Standard Model charm $CP$ violation in the 
interference of decay and mixing, while not only observing but characterising $CP$ violation in decay across 
a broad range of charm hadrons and decays. These capabilities will remain unique for the foreseeable future 
unless a dedicated fixed-target experiment, such as the proposed TauFV charm factory, is built at the 
LHC, which seems unlikely at present.

Taken together, the capabilities of FCC-ee, Belle~2, and LHCb Upgrade~2 provide a powerful motivation to continue
the exploration of quark flavour for the next decades. Sensitivities to exotic hadrons are harder to quantify.
It is however clear, particularly in light of LHCb's demonstration that a triggerless flavour factory can be
operated at a hadron collider, that their datasets (as well as the complementary and more focused ones of BESIII) 
will leave a unique legacy of sensitivity to all possible types of hadrons and their decays. These capabilities
will be complemented by an ultimate exploration of the kaon sector by NA62. If an experiment can be built to
precisely study the $K^0\to \pi^0 \nu \bar{\nu}$ decay, this result and NA62's measurement of 
$K^+\to \pi^+ \nu \bar{\nu}$ will give yet another fully independent measurement of the apex of the unitarity
triangle. The only frustration is that such a programme will, on the most optimistic projections, take
another forty-odd years to execute, requiring not only tremendous commitment from the current generation
of experimentalists but convincing many new generations to join what is a fundamentally incremental rather
than revolutionary effort. Nevertheless experience with the construction of the current, upgraded, LHCb
detector and the construction and operation of Belle~2 clearly demonstrate that these timescales are not
caused by lack of effort or willpower but rather by the fundamental difficulty of the problem we are trying
to solve. Perseverence, and a transparent communication of both the challenges and motivations for this effort,
must therefore be the order of the day.

\section{Conclusion}

It is difficult not to feel dizzy when contemplating the progress made in understanding the most
fundamental constituents of nature over the past hundred and thirty odd years. A physicist studying
at the end of the nineteenth century wouldn't have even been able to take the structure of an atom
as a given, nevermind contemplate an internally coherent and self-consistent theory of all microscopic
reality which has withstood decades of concerted tests by tens of thousands of physicists. It is easy
to imagine this rate of progress will continue and allow us to complete the Standard Model by finding
additional particles and interactions which reconcile it with cosmology and, why not, gravitation.
As bullish stock market investors are fond of saying, the best time to buy shares in beyond Standard 
Model searches was yesterday and the second-best time is today.

And yet, scientific knowledge does not progress in a linear fashion. Bursts of progress in our conceptual 
understanding are often punctuated by long plateaus (and sometimes dips) during which factual knowledge 
is accumulated but we cannot see the reference frame in which these pieces of the puzzle fit together. 
More than two and a half millenia divide initial recorded ideas about atoms as fundamental constituents 
of nature from the miraculous explosion of understanding during the twentieth century which
culminated in the Standard Model. Hundreds of years separated the original Copernican model from
the general theory of relativity. There is no guarantee that we have not reached another such plateau 
today, nor of how long it might last.

On the contrary, quark flavour physics provides a fair amount of evidence that such a plateau may stretch 
out before us. The hierarchical structure of the CKM matrix is a puzzling feature of the Standard
Model, not mandated by any fundamental principles or symmetries. Generic models of new particles or
interactions which couple to quarks therefore do so with order one coupling strengths, and are already
ruled out at tens of thousands of TeV. So new physics must either be beyond the reach of any conceivable
laboratory experiment, not couple to quarks, or do so with highly suppressed couplings. Of course as
experimentalists we focus on scenarios we can actually test, which amounts to searching for new physics
with suppressed couplings to quarks (and possibly couplings which arise only at loop-level). It is
correct and necessary to push existing technology to its limit and search in the most exhaustive manner
for new physics -- a quest in which quark flavour physics will continue to complement direct searches
as described in the preceeding sections. But if that were all quark flavour physics could do, it would
also not be unreasonable to question how much human effort should really be dedicated to what increasingly
looks like a cause grounded more in hope than expectation.

Fortunately, quark flavour physics remains far more than a vehicle for trading futures in beyond 
Standard Model particles and interactions. There is a tremendous amount which we do not yet understand
about interactions whose existence is not in doubt, most notably the strong force, where quark flavour
physics continues to drive both factual and conceptual progress. As for the weak force, a few-permille 
metrology of the CKM matrix is within grasp, and this knowledge will be crucial to characterise new 
physics if and when it is eventually found. It is easy to think of these 
objectives as a byproduct of more exciting endeavours, but that would be a serious mistake. Decades of 
painstaking experimental work remain before us if they are to be achieved, and this will require not 
only focus from the current generation of high-energy physicists. It will also require a willingness 
to educate and enthuse the next generations in the fundamental importance of this quest in and of itself.

Perhaps this quest will result in a revolutionary revelation of beyond SM physics after all. In this case,
the tools we have built to find what lies beyond the Standard Model will also serve us well to characterise
its properties. But if it does not, the coming decades might well see the final generation of collider 
experiments and facilities which are built, broadly speaking, in the image of the SPS and its UA1 and UA2 
detectors. The available room for increases in energy or instantaneous luminosity becomes ever smaller, while 
technologies with a potential to break out of this plateau, such as the muon collider, still seem some way from 
being viable. If this is to be the end of an undoubtedly golden era in humanity's quest for a fundamental 
understanding of nature, then it remains for those of us who came of age in its twilight to ensure that we 
leave behind the most expansive and accurate legacy of knowledge possible to future generations. This means 
refusing to compromise on the importance of our fundamental measurements and pushing technology to its utmost 
limits to build the most precise, accurate, and versatile detectors to deliver this legacy. Even if it takes 
one hundred years one day a new generation will have the tools to go further, and 
the knowledge we accumulated will be their companion on that journey. That is its own reward.

\section*{Acknowledgments}
I thank colleagues from the LHCb, Belle~2, BESIII, and UTFit collaborations for their generous help in preparing 
much of the material which went into this review. I would also like to express my appreciation for the work 
of the PDG as well as the HFLAV and CKMFitter averaging groups whose figures and results were an equally crucial 
part of this review. I thank M\'{e}ril Reboud, Francesco Dettori, Dean Robinson, and Vincenzo Vagnoni for stimulating 
comments. I also thank M\'{e}ril Reboud and for producing an updated global
fit of the $b\to s\mu^+\mu^-$ processes for this review. I thank Carla G\"{o}bel for a careful proofreading of 
the manuscript and helpful comments. This work is supported by the European Research Council 
under Grant Agreement number 724777 ``RECEPT''. 

\bibliographystyle{ws-ijmpa}
\bibliography{my-bib-database}
\end{document}